\documentclass{mn2e}
\usepackage{graphics}
\def\hi{H\,{\sc i} }
\def\hii{H\,{\sc ii} }

\def\kmss{km~s$^{-1}$ }
\def\kms{km~s$^{-1}$}

\def\dispr{{\langle}v^2_R{\rangle}^{1/2}}

\def\dispt{{\langle}v^2_{\Theta}{\rangle}^{1/2}}
\def\dispz{{\langle}v^2_z{\rangle}^{1/2}}
\def\dispzn{{\langle}v^2_z{\rangle}^{1/2}_{R=0}}
\def\msol{${\rm M}_{\sun}$}
\def\smppc2{${\rm M}_{\sun} {\rm pc}^{-2}$}

\def\epssn{{\varepsilon}_{\rm SN}}
\def\vmd{v^{\rm max}_{\rm disc}}
\def\vmo{v^{\rm max}_{\rm obs}}
\def\msolperyr{M_{\sun} {\rm yr}^{-1}}
\begin{document}
%\thesaurus{03(11.09.1 NGC 3992;  09.11.1 11.06.2 11.11.1 11.19.2)} 
\title[Simulations of normal spiral galaxies]{Simulations of normal spiral galaxies}
%\subtitle{  }
\author[Roelof Bottema]{Roelof Bottema\\
Kapteyn Astronomical Institute, P.O. Box 800, 
NL-9700 AV Groningen, The Netherlands, e-mail:robot@astro.rug.nl}
%\date{Received date1; accepted date2}
%
%
%
\maketitle
\begin{abstract}
Results are presented of numerical simulations of normal isolated
late type spiral galaxies. Specifically the galaxy NGC 628 is used
as a template. 
The method employs a TREESPH code including
stellar particles, gas particles, cooling and heating of the gas,
star formation according to a Jeans criterion, and Supernova feedback.
A regular spiral disc can be generated as an equilibrium situation
of two opposing actions. On the one hand cooling and dissipation of the
gas, on the other hand gas heating by the FUV field of young stars and 
SN mechanical forcing.
The disc exhibits small and medium scale spiral structure of which 
the multiplicity increases as a function of radius.
The theory of swing amplification can explain,
both qualitatively and quantitatively,
the emerging spiral structure. In addition, swing amplification
predicts that the existence of a grand design $m=2$ spiral
is only possible if the disc is massive.
The simulations show that the galaxy is then unstable to bar 
formation, confirming the result of Ostriker \& Peebles (1973).
The occurrence of this bar instability is further investigated.
A general criterion is derived for the transition between bar
stable and unstable, depending on disc mass contribution
and on disc thickness. 
It seems that bar stability hardly depends on the presence of gas. 
A detailed quantitative analysis is made of the emerging spiral structure
and a comparison is made with observations. That demonstrates that
the structure of the numerical isolated galaxies is not as strong
and has a larger multiplicity compared to the structure of some
exemplary real galaxies. It is argued that 
the suggestion of Kormendy \& Norman (1979) holds, that
a grand design can only
be generated by a central bar or by tidal forces resulting from an
encounter with another galaxy.
\end{abstract} 
\section{Introduction}
It is now well established that substantial amounts of dark matter
are needed to explain flat rotation curves in the outer regions
of spiral galaxies. However, the amount of dark matter present in
the inner, luminous region of a galaxy is not well determined.
Does a spiral galaxy conform to the maximum disc hypothesis
(van Albada \& Sancisi 1986; Salucci et al. 1991; Sellwood \& Moore 1999)
which states that the luminous part should be scaled up as much as possible?
Or is the maximum rotational contribution of the disc to the 
total rotation around 63\%? The latter being derived from observations
of disc stellar velocity dispersions (Bottema 1993, 1997) or suggested
by a statistical analysis of rotation curve shapes in relation
to the compactness of discs (Courteau \& Rix 1999).
When discs obey the 63\% criterion, their mass-to-light ratio is 
half that of the maximum disc case, but the disc still dominates
in the inner two to three scalelengths. A rotation curve analysis
generally allows a maximum disc rotational contribution anywhere
between 50 to 90\% (van der Kruit 1995). The above mentioned velocity
dispersion measurements have only been done for a dozen galaxies and 
the analysis depends on the not well determined ratio of disc
scalelength to scaleheight. As observations are not yet
conclusive, numerical simulations of spiral galaxies may give clues
regarding the disc to halo mass ratio.

Spiral structure of galaxies has eluded astronomers already for
a long time. Is it a quasi stationary density wave (Lin \& Shu 1964,
1966)? Or is it a temporal phenomenon excited by internal
or external disturbances and then swing amplified (Goldreich \&
Lynden-Bell 1965; Toomre 1981)? Or is it some kind of combination, 
like swing amplified density waves? Until now theory and observations
are inconclusive. To compound matters, there is the ``anti spiral
theorem'' (Lynden-Bell \& Ostriker 1967), which states that spiral
structure cannot develop in a system governed by a time reversible
description. Consequently spiral arms cannot exist in an isolated
pure stellar galaxy. Some kind of non time reversible action is needed,
like external forcing or gas dissipation. The need for the latter
mechanism is confirmed observationally; in general when there is 
no gas there are no spiral arms. When studying spiral structure one
thus has to include a gaseous component in the calculations. Doing
that, calculations in three dimensions are required because for a
typical disc the scaleheight of the stellar component is five
times as large as the scaleheight of the gas layer.

Some galaxies have bars, others don't.
Ostriker \& Peebles (1973) showed that a stellar disc
can be stabilized against bar formation by adding a spherical
(dark halo) potential. This has lead to the common believe that
barred galaxies have a larger disc to halo mass ratio compared to
non-barred galaxies, although there is no observational evidence
to support this believe. Other theoretical and numerical calculations
(Efstathiou et al. 1982; Bottema \& Gerritsen 1997) investigating this
matter reconfirm that a relatively larger disc mass results in a galaxy which
is more unstable to bars. Therefore this ratio is inevitably an important
parameter in determining the appearance of a galaxy. Another
effective parameter has to be the stellar velocity dispersion
of a disc. Larger dispersions, expressed in a larger Toomre's (1964)
$Q$ value, lead to a more stable disc (Sellwood \& Carlberg 1984).
In addition a substantial bulge makes a disc more stable against bars
(Sellwood \& Evans 2001).

A proper investigation of disc galaxies involves a.o. dissipational
and non linear processes. Therefore this subject is not easily
handled by analytical calculations and one has to resort to numerical
simulations. Results of such simulations can then guide the way
to more detailed analyses of some specific aspects. Concerning numerical
simulations, the combination of the tree code (Barnes \& Hut 1986)
for the collisionless part of a galaxy with the Smoothed Particle
Hydrodynamics (SPH) algorithm (Lucy 1977) for the collisional part allows
the study of a galaxy containing gas. However, this combined
TREESPH code (Hernquist \& Katz 1989) has rarely been used
to study the detailed evolution and stability of isolated galaxies.
It has mainly been applied to investigate interacting galaxies
(Barnes \& Hernquist 1991; Mihos \& Hernquist 1994; Springel 2000),
galaxy formation (Katz \& Gunn 1991; Katz 1992) or barred galaxies
specifically (Friedli \& Benz 1993, 1995).
A preliminary numerical study of spiral structure
has been performed by Elmegreen \& Thomasson (1993) while 
Gerritsen (1997) and Gerritsen \& Icke (1997, hereafter: GI97) made a detailed
numerical analysis of the distribution and 
interaction of the stellar and gaseous components
in an isolated galaxy. 

For a galaxy simulation one needs at least stars and gas.
An optional bulge can be added and in order to make the observed flat
rotation curves, a dark halo should be included.
When gas is present, that cools and dissipates so that a star
formation (SF) prescription is needed. That leads to gas depletion
and to the formation of young stars with low velocity dispersions. 
SF also generates heating and stirring of the gas, preventing
it from further collapse and so establishing a feedback loop which
creates some kind of interstellar gas medium in equilibrium. These
processes are too complex to be simulated from basic principles
so that one needs a global description
for the entities and processes in galaxy simulations. 

Concerning
SF there is a clear observational result, namely the Schmidt (1959) law
which states that the SF rate (SFR) is proportional to the gas density
to a power $n$: SFR $\propto {\rho}^n$. Detailed observations and analyses
by Kennicutt (1983, 1989, 1998) and by Ryder \& Dopita (1994)
show that this Schmidt law holds 
with $n \sim 1.3$ for a wide range of (surface) densities in galactic discs.
Consequently this observation and proper exponent has to be the outcome of 
any numerical SF prescription. The interesting result of GI97
is that they reproduce the right Schmidt law by using a star
formation law based on a Jeans criterion. They do not impose the Schmidt
law a priori, but it is the result of a more fundamental process.
Their scheme is therefore expected to hold for a broad range
of physical conditions. 

It is well known that in a real galaxy young stars and SF regions
exert a definite action on the ambient ISM. The fierce radiation
heats the gas and supernova (SN) action stirs up the ensemble of
gas clouds. This SF feedback maintains an equilibrium state of the ISM.
When simulations lack such a feedback a catastrophic gas cooling
occurs, forming super giant cloud complexes (Shlosman \& Noguchi 1993).
Therefore in the numerical calculations also a SF feedback prescription
is needed.

As mentioned above, Elmegreen \& Thomasson (1993) made a numerical
study of isolated galaxies to specifically investigate spiral
structure, more or less as a function of disc to halo mass ratio.
They find for a ``rather massive'' disc a grand design $m=2$ structure
which lasts for several rotation periods. For substantially less massive
discs the spiral structure becomes flocculent. Though their
investigation is very interesting it contains several caveats.
The main one is that the development of a bar is prevented in at 
least two ways. Firstly by using an artificial $Q$ barrier in the
inner regions. Secondly the disc is stabilized by a large softening
length (Romeo 1994) which has to be used to perform the two dimensional
simulations. Therefore it is expected 
that the results of Elmegreen \& Thomasson only partially apply
to a real galaxy. Results in the present paper differ from the
study of Elmegreen \& Thomasson mainly relating to the development
of a bar in massive discs.

My original intention was to investigate what happens to the morphology
of a normal galaxy when the ratio of disc to dark matter changes. 
Inevitably, during the research other and additional problems emerged
which necessitated a somewhat broader scope. Nevertheless this paper
still focuses on the global properties and global morphology
of a galaxy and does not aim to go into specific details.
Besides the abovementioned mass ratio another defining parameter for
a galactic disc is the value of the stellar velocity dispersion, or
more or less equivalently, the thickness of the disc. In addition
effects of the amount of SN feedback and amount of available gas have 
been considered. The investigations are restricted to late type
galaxies for which NGC 628 serves as template. That galaxy is nearly
face-on, has only a minor bulge, and has a well defined, non barred 
spiral structure.

Section 2 of this paper describes the code and all the relevant
prescriptions. The employed numerical scheme is based on the works
of GI97, Gerritsen (1997), and Hernquist \& Katz (1989). Besides
the nice asset of reproducing the Schmidt law, the scheme creates
a multiple phase interstellar medium, equal to the situation in a
real galaxy. GI97 performed their simulations for galaxies with low
SF rates and then an equilibrium gas state can be achieved by UV
heating only. When the SFR crosses a certain threshold it is 
necessary to include mechanical feedback in addition. This supernova
feedback will be desribed in sufficient detail.
In Sect. 3 the model galaxy is described and in
Sect. 4 several properties and adaptations are discussed. To create
a realistic disc in star formation equilibrium in Sect. 5 the necessary
amount of SN feedback and external gas supply are determined. Section 6
describes what happens when the main parameters; disc to halo mass
ratio and disc thickness are varied. When does a bar form and what is the
global morphology? In Sect. 7 an analysis and discussion is given
of the gas properties, like temperature distribution and gas velocity
dispersion. Section 8 deals in some detail with the spiral structure
and compares results from the simulations with observations.
In Sect. 9 the theory of swing amplification is applied to the
galaxies under investigation and finally in Sect. 10 
a general discussion is given and a point by point
summary of the conclusions. 
\section{The code}
\subsection{Numerical technique}
The evolution of a galaxy is simulated using a hybrid N-body/hydrodynamics
code (TREESPH; Hernquist \& Katz 1989). A tree algorithm
(Barnes \& Hut 1986; Hernquist 1987) determines the gravitational 
forces on both the stellar and gaseous components of the galaxy.
The hydrodynamic properties of the gas are modelled using the
method of smoothed particle hydrodynamics
(Lucy 1977; Gingold \& Monaghan 1977). The gas evolves according
to hydrodynamic conservation laws, including an artificial viscosity
for the treatment of shocks. Each gas particle is assigned an
individual smoothing length, $h_i$, which determines the
local resolution. Gas properties are found by smoothing over $N$
neighbours within $2h$, where
$N$ is typically 64 for a number of gas
particles larger than 12.000. The acceleration $(\vec{a_i})$
of particle $i$ is given by
\begin{equation}
\vec{a_i} = - \frac{1}{{\rho}_i} \nabla P_i + {\vec{a_i}}^{\rm visc}
- \nabla {\Phi}_i ,
\end{equation}
with $P$ being the pressure, ${\vec{a_i}}^{\rm visc}$
the artificial viscosity term (Monaghan 1989), and $\Phi$ the gravitational
potential. For the evolution of the specific thermal
energy $(u)$ an expression is chosen as
\begin{equation}
\frac{du_i}{dt} = \frac{1}{2} \vec{v_i} \cdot
(\frac{1}{{\rho}_i} \nabla P_i - {\vec{a_i}}^{\rm visc} )
+ \frac{\Gamma - \Lambda}{\rho},
\end{equation}
where $(\Gamma - \Lambda)/{\rho}$ accounts for the non-adiabatic
heating and cooling terms. To close the system of equations
describing the evolution of the fluid an equation of state
\begin{equation}
P = (\gamma - 1) \rho u,
\end{equation}
is adopted with $\gamma = 5/3$ for an ideal gas.

Time steps of the stellar particles were equal to 
1.5~10$^6$ years. For
gas (SPH) particles a variable time step was used determined by
the Courant-Friedrichs-Levy condition, with a maximum
of 1.5~10$^6$ years. The softening of stellar particles
was not implemented by the usual softened force law: 
$a \propto (r^2 + {\varepsilon}^2 )^{-1}$ but by
a smoothing kernel surrounding the stellar particle
(Hernquist \& Katz 1989).
In that way a better approximation is achieved of the real
force law.

\subsection{The cooling}
The cooling term $\Lambda$ in Eq. (2) describes the radiative
cooling of the gas. For the simulations a constant composition of
\hi plus Helium gas is adopted. The ``standard'' cooling
function of Dalgarno \& McCray (1972) was used for
this gas, parameterized as
\begin{eqnarray}
\frac{\Lambda}{[{\rm erg}\;{\rm cm}^{-3}\;{\rm s}^{-1}]} & = &
10^{-21} n_H^2 {\Bigl[} 10^{-0.1 - 1.88(5.23 - {\rm log}T)^4} +
\nonumber \\
& &10^{-a - b(4 - {\rm log}T)^c} {\Bigr]} \;\;\;
{\rm if}\;\; {\rm log}T < 6.2
\end{eqnarray}
and
\begin{equation}
\frac{\Lambda}{[{\rm erg}\;{\rm cm}^{-3}\;{\rm s}^{-1}]} =
10^{-22.7} n_H^2 \;\;\; {\rm if}\;\; {\rm log}T > 6.2 ,
\end{equation}
where $T$ is the temperature and $n_H$ the hydrogen density
in cm$^{-3}$. The second term on the right hand side of Eq. (4)
determines the cooling below 10$^4$~K, which is strongly dependent
on the amount of ionization of the \hi gas. For an ionization 
parameter $n_e/n_H$ of 0.1 (Cox 1990) a good approximation
of Dalgarno \& McCray's cooling function is found
for $a = 3.24,\; b = 0.085,$ and $c = 3.0$.

The star formation rate is dependent
on the amount of cooling below 10$^4$~K and the parameter $a$
mainly determines this amount. For a lower ionization or for
lower abundances the cooling will be smaller and the SFR lower
(Gerritsen \& de Blok 1999). In the calculations the cooling
rate for all gas is set a priori equal to approximately the solar
neighbourhood value. The effects of metallicity changes in a galaxy have not
been considered.

\subsection{Heating}
The heating of the gas $(\Gamma)$ is described by two terms:
\begin{equation}
\frac{\Gamma}{[{\rm erg}\; {\rm s}^{-1}\; {\rm cm}^{-3}]}
= 10^{-24} \epsilon G_0 n_H + C_{\rm cr} n_H ,
\end{equation}
where the first term represents photoelectric heating
of small grains and PAHs (de Jong 1977; Wolfire et al. 1995)
and dominates in the stellar disc region. 
In Eq. (6) $\epsilon$ is the heating efficiency and $G_0$ is the
incident far-ultraviolet field (912 to 2100 \AA)
normalized to Habing's (1968) estimate of the
local interstellar value (= 1.6 10$^{-3}$ erg cm$^{-2}$ s$^{-1}$).
This heating efficiency $\epsilon \approx 0.05$ for temperatures
below 10.000~K. The
second term in Eq. (6) represents the so called 
``cosmic ray'' heating (Black 1987; Spitzer 1978).
A value of 7.8~10$^{-27}$ erg~s$^{-1}$ has been used for the cosmic
ray heating constant $C_{\rm cr}$. In a normal
stellar environment cosmic ray heating is negligible compared
to the grain heating. Yet in the outer regions of a galaxy,
beyond the stellar edge, it is still small, but it is
the only heating source. This term has been included after the work
of GI97 to prevent too much cooling and star formation in those
outer regions.

Since the gas is heated mainly by FUV photons, the stellar
radiation field has to be calculated. In the simulations each
stellar particle effectively represents a stellar association. 
Such an association is formed instantaneously and consequently
each stellar particle has an age and a belonging FUV flux according
to that age. This flux has been calculated using the population
synthesis models of Bruzual \& Charlot (1993) for a certain
specified IMF. Knowing
the FUV flux for every stellar particle, the radiation
field is calculated by summing the individual fluxes corrected
for geometric dilution. No extinction corrections have been
applied.
The cooling of the gas increases by more than a factor 10$^3$
for temperatures rising above 10$^4$~K. That temperature
is therefore effectively an upper limit, virtually independent
of the heat input.

\subsection{The star formation criterion}

Star formation on scales $\la$ giant molecular clouds is a very
complicated process. It involves matters like heating/cooling balance,
collapse, fractionization and sub fractionization, the transition
of \hi into H$_2$, magnetic fields,
winds, SN explosions, shocks, ionization, desintegration, etc, etc.
GMCs, where the star formation takes place, have typical masses of
a few times 10$^5$ to 10$^6$ $M_{\sun}$, densities $>$ 10$^3$
cm$^{-3}$ and lifetimes of 10$^7$ to 10$^8$ years (Shu et al. 1987;
Bodenheimer 1992). Masses of gas particles in the simulation range
typically from 5~10$^5$ to 10$^6$ solar masses, being comparable
to GMCs. A star formation recipe is thus a prescription of converting
GMCs into stellar associations and does not consider all the 
processes mentioned above.

The formation of stellar associations ($\equiv$ stellar particles)
proceeds nearly equal as described by GI97. For
convenience a summary of this star formation 
process will be given. In SPH simulations the gas particles sample
properties such as density and temperature. For each particle
one can therefore calculate the Jeans mass $M_J$
\begin{equation}
M_J = \frac{1}{6} \pi \rho \left( \frac{ \pi c_s^2}{G \rho}
\right)^{3/2},
\end{equation}
where $\rho$ is the density, $c_s$ the sound speed and $G$ the
gravitational constant. If this Jeans mass is smaller than a
prescribed critical mass $M_{\rm crit}$, which is of the order
of the mass of a GMC, then the gas particle is in a cold and dense
environment and is considered unstable to star formation. So,
the condition $M_J < M_{\rm crit}$ is a first criterion a gas
particle has to obey to form a stellar particle.

Once a region is unstable, it takes a collapse time to form a stellar
cluster. As a second criterion it has been chosen that a gas
particle has to remain unstable for a time span longer than
a fraction $f_c$ of the free-fall time $t_{ff}$:
\begin{equation}
t_{\rm span} > f_c \times t_{ff} = \frac{f_c}{\sqrt{4 \pi G \rho}}.
\end{equation}
As soon as an SPH particle has fulfilled both conditions (7) and (8)
a part of its mass ($M_g$) is converted into a stellar particle
with mass $M_{\ast}$
\begin{equation}
M_{\ast} = {\epsilon}_{\rm sf} \times M_g.
\end{equation}
This continues untill the mass of a gas particle would drop below
0.1 times its original mass, then the whole remaining gas particle
is turned into a new star.
The star particle is given the same IMF as all other
star particles, and age zero. New star particles are then included
in the calculation of the radiation field and the ambient gas will
be heated. In this way a self regulating star formation process
is established. 

GI97 made a thorough investigation of the effects when a specified
SF parameter is changed. In short, the cooling function essentially
determines the SFR and a change of the collapse factor $f_c$ changes 
the ratio of warm to cold gas. Changing other parameters has a
negligible or only a small effect on observable structures and quantities.
In the present study the star formation parameters are essentially
fixed at a certain value. The critical mass $M_{\rm crit}$
is equal to the value of a GMC, the IMF is Salpeter (0.1 $\rightarrow$
125 $M_{\sun}$), the cooling function is solar neighbourhood and
${\epsilon}_{\rm sf} = 0.5$. The collapse factor was determined
at 0.5 such that the gas is not too clumpy and the warm to cold 
gas ratio is reasonable (see Sect. 5.1 and 7.2).

%Fig1
\begin{figure}
\resizebox{\hsize}{!}{\includegraphics{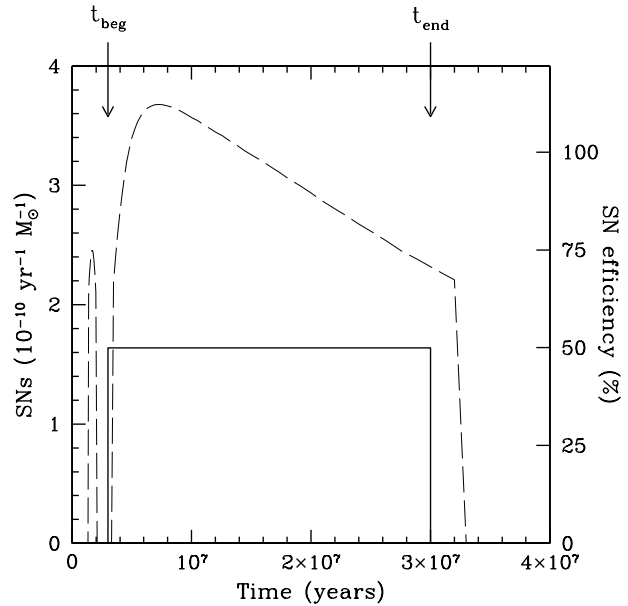}}
\caption[]{Graphical representation of supernova (SN) feedback.
The dashed line and left axis give the number of SNs as a function of time
following Bruzual and Charlot (1993). In the code 
the mechanical SN feedback is 
implemented by giving a certain fraction (SN efficiency) of
the belonging SN energy as a constant thermal energy to one SN particle.
This is illustrated by full drawn line and right axis
for a SN action from $t = t_{\rm beg}$ to $t = t_{\rm end}$
for $\epssn$ = 50\%.}
\end{figure}

\subsection{Supernova feedback}

Besides the FUV feedback on the gaseous medium there is also
the mechanical feedback by supernovae and stellar winds.
A method to implement that feedback has been developed by
Gerritsen (1997) which I have largely taken over. Since a description
has not appeared in the refereed literature and because a number
of changes have been made to the method, a detailed description
now follows.

Winds and supernovae from OB associations are capable of blowing
cavities in the surrounding ISM. In the standard model for such
a process the mechanical energy is injected into a hot bubble which
sweeps up the external medium by a dense shell (McCray \& Kafatos 1987).
Not all the energy is converted into kinetic energy of the expanding
shell. Silich et al. (1996) demonstrate that most of the energy
is simply radiated away and only 5 to 10\% ends up in the expanding
shell. An amount of $\sim$ 10$^{51}$ ergs of energy is provided
by each supernova so that every star more massive than 8~$M_{\sun}$
supplies that amount to the mechanical feedback process. Additional
energy is provided by winds from massive and medium size stars
(Lamers \& Leitherer 1993).

The more classical implementation of mechanical feedback is by
``kicking'' the surrounding ISM. The gas around a young star cluster
is then given a velocity perturbation directed radially away from
the cluster. A free parameter in this implementation is the fraction
of the released energy $(f_v)$ that is put into the velocity
perturbation. The feedback may have profound effects on the gas
depending largely on the value of $f_v$. For example, Navarro 
\& White (1993) favour values of $f_v \la 0.1$ in their cosmological
simulations. But Friedly \& Benz (1995) take of the order of 0.01
for barred galaxies and Mihos \& Hernquist (1994, 1996) use 
$f_v \sim 10^{-4}$ for simulations of interacting galaxies. Clearly
there is a lot of arbitraryness as to the choice of $f_v$.

At present a different approach to implement mechanical feedback
has been chosen. Following an idea by Heller \& Shlosman (1994)
a fraction of the mechanical energy is given as thermal energy to one
gas particle during a certain amount of time. In this way a hot bubble
is simulated by making one SPH particle hot; from now on called
SN particle. It is further left to the TREESPH program to
distribute the energy to the surrounding gas particles like a
pseudo-point explosion. 

For the implementation in the code it has to be realized that one is
dealing with particles representing GMCs and whole star forming
regions. When a gas particle (GMC) is deemed to form a star particle
(SF region) half of its mass is transformed into a star particle
with zero age and belonging IMF. The other half of the gas particle
is designated as SN particle possessing the mechanical energy appropriate
for the SF region. For a period of $\sim$ 3~10$^7$ years SN activity
is present in the SF region and for such a time the SN particle is both,
fixed to the parent star particle, and kept hot. To illustrate matters,
in Fig. 1 the SN energy is given as a function of time according
to the Bruzual \& Charlot population synthesis code. On top of that
figure the simplified SN particle energy is displayed. For a small
time span after star formation SN activity is not yet switched on
and the SN particle is kept at a temperature of 10$^4$~K. After that
a fraction $\epssn$ of the total SN energy of the SF region is given as
thermal energy to the SN particle. This total SN energy is calculated 
from the total mass of the new star cluster, the assumed IMF, and
the fact that all stars go SN for a mass $>$ 8~$M_{\sun}$. 
After a time $t_{\rm SN}^{\rm end}$ no more
energy is supplied to the SN particle and it is allowed to cool again
which it will do quickly in practice. 

In reality there is a time variability of the SN action (Fig. 1) 
and a poorly known energy contribution from winds. These details
and uncertainties have been included in the free parameter $\epssn$.
After the SN phase
the gas particle returns to being a normal SPH particle which may
undergo a next event of star formation if the conditions are right.
Most simulations in this paper are performed for $\epssn =$ 25\%,
and for a regular SF where half the gas mass is transformed into stars,
the belonging temperature of the SN particle amounts to 1.25~10$^7$~K.
In this, albeit artificial, way a three phase ISM is created containing
cold, warm, and hot gas (McKee \& Ostriker 1977).

Gerritsen (1997) did some testing of this method on isolated galaxies.
It appeared that realistic cavities are created in the ISM, where the
holes are smaller in the central denser regions and larger near the rim
of the gas distribution. The size and the size distribution of the holes
does not depend on the number of gas particles. This demonstrates
the reliability of the method and shows it can be used for
a large gas density range even at a limited resolution.
Compared to a situation with only FUV heating of the gas, adding
the mechanical feedback
changes the gas appearance to a different and larger design.

\section{The galaxy}
All the calculations have been performed for one model galaxy with
different internal compositions. This galaxy consists of four
components: a stellar disc, a gas layer, a bulge, and a dark halo.
This disc and gas layer are made up of stellar and gas particles while
the bulge and halo are implemented as rigid, ``dead'', potentials.

%Fig2
\begin{figure}
\resizebox{\hsize}{!}{\includegraphics{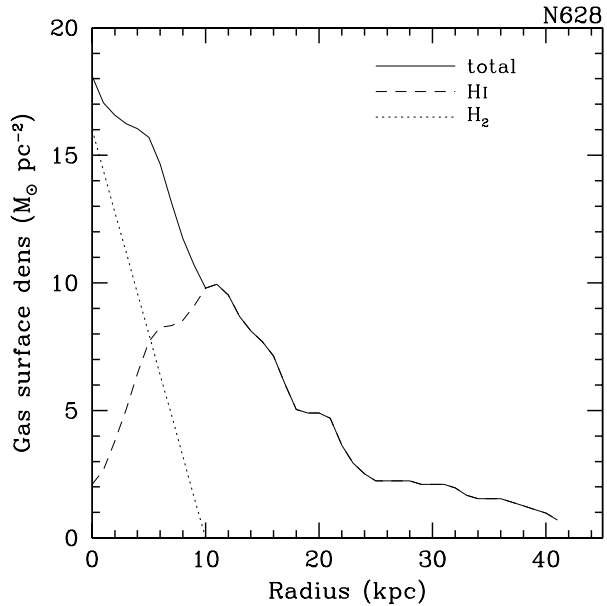}}
\caption[]{The total gas surface density of NGC 628
which has been used in the model galaxy. \hi and H$_2$ have been
added and multiplied by 1.4 to account for Helium.}
\end{figure}

An exponential stellar disc is assumed with a locally
isothermal z-distribution (van der Kruit \& Searle 1981)
\begin{equation}
{\rho}_d(R,z) = {\rho}_0^d {\rm e}^{-R/h} {\rm sech}^2
\left( \frac{z}{z_0} \right),
\end{equation}
where $h$ is the scalelength of the disc and $z_0$ the thickness
parameter which was constant as a function of radius. If the disc
were isolated the vertical stellar velocity dispersion $\dispz$
depends on the stellar surface density ${\sigma}_{\ast}$ as
\begin{equation}
\dispz = \sqrt{ \pi G z_0 {\sigma}_{\ast}(R) }.
\end{equation}
In reality adjustments are necessary to this relation because
the other galactic components influence the stellar velocity dispersion.
For the setup of the galaxy only the gas has been taken into account
such that in Eq. (11) ${\sigma}_{\ast}(R)$ is replaced by
${\sigma}_{\ast}(R) + {\sigma}_{\rm gas}(R)$. During the initial
stages of the simulations the stellar disc still has
to settle in the bulge and halo potential, but in general such a
disc adaptation is small, certainly if the disc is relatively
massive. The dispersions in the radial and tangential directions are
related to $\dispz$ as
%
%
%TABLE 1
\begin{table*}
\caption[]{Galaxy parameters}
\begin{flushleft}
\begin{tabular}{lllllllll}
\hline
\noalign{\smallskip}
Disc & Disc & Disc central & $\dispzn$ for & Disc &
Halo & Halo & Bulge & Bulge central \\
$v_{\rm max}$ & mass & surf. dens. & $z_0$ = 900 pc & $(M/L_B)$ &
$R_{\rm core}$ & $v_{\rm max}$ & mass & surf. dens. \\
(\kms) & (10$^9$ \msol ) & ({\smppc2}) & (\kms) & &
(kpc) & (\kms) & (10$^9$ \msol) & ({\smppc2}) \\
\noalign{\smallskip}
\hline
\noalign{\smallskip}
126 & 45.5 & 367 & 67 & 2.05 & 3    & 185 & 2.4 & 2562 \\
150 & 64.5 & 520 & 80 & 2.90 & 5    & 180 & 3.4 & 3630 \\
165 & 78.0 & 629 & 88 & 3.51 & 11.2 & 207 & 4.2 & 4390 \\
180 & 92.9 & 749 & 96 & 4.18 & 23.4 & 264 & 5.0 & 5230 \\
\noalign{\smallskip}
\hline
\noalign{\smallskip}
\multicolumn{9}{l}{\quad Distance = 11.5 Mpc} \\
\multicolumn{9}{l}{\quad Total light in B = 22.2 10$^9$ $L^B_{\sun}$ (uncor.)} \\
\multicolumn{9}{l}{\quad Total gas mass = 18.8 10$^9$ \msol }\\
\multicolumn{9}{l}{\quad Disc scalelength = 4.5 kpc}\\
\multicolumn{9}{l}{\quad Bulge scalelength = 397 pc}\\
\multicolumn{9}{l}{\quad Disc mass / bulge mass = 18.7}\\
\end{tabular}
\end{flushleft}
\end{table*}
\begin{equation}
\dispz = 0.6\; g(R) \dispr,
\end{equation}
and
\begin{equation}
\dispt = \frac{\kappa}{2\Omega} \dispr .
\end{equation}
Here $g(R)$ is a function close to 1.0 and will be discussed in the
next section. The tangential dispersion $\dispt$ is related to the
radial dispersion through $\Omega$ and $\kappa$, which are the
orbital and epicyclic frequencies. The average rotation of the
stars $(v_{\ast})$ is given as the testparticle rotation,
$v_t = \sqrt{(d\Phi / dR)/R}$ diminished with the asymmetric drift:
\begin{equation}
v_t^2 - v_{\ast}^2 =
\langle v^2_R \rangle \left[ \frac{R}{h} - R \frac{\partial}{\partial R}
{\rm ln}\langle v^2_R \rangle + \frac{1}{2} \left(
\frac{R}{v_{\ast}} \frac{\partial v_{\ast}}{\partial R} -1 \right) \right],
\end{equation}
using a plane parallel epicyclic approximation. At the radii where the
stellar velocity dispersion becomes larger than the rotation, Eq. (14)
does not hold any more. For these inner regions an interpolation
by hand has been made to $R = 0$. Stellar particles are distributed
according to Eq. (10) and are given a rotation according to Eq. (14).
Dispersions in the $R, z,$ and $\Theta$ directions follow from
Eqs (11) to (13) and are drawn from Gaussian distributions.

The gas layer has a radial functionality as observed for NGC 628. The
thickness of the gas layer is determined by the surrounding potentials
of the other components and the gas velocity dispersion.
For the latter a constant, isotropic value of 8 \kmss has been adopted.
To calculate the thickness of the gas layer the (ad hoc) description
of Bottema (1996) has been used.

The bulge is spherical with a projected exponential radial distribution,
${\sigma}_{\rm b} = {\sigma}^{\rm b}_0 \; {\rm e}^{-r/h_{\rm b}}$ resulting
in a spatial distribution of the bulge density ${\rho}_{\rm b}$ of
\begin{equation}
{\rho}_{\rm b} = \frac{ {\sigma}_0^{\rm b}}{\pi h_{\rm b}}
K_0 \left( \frac{R}{h_{\rm b}} \right).
\end{equation}
Here $h_{\rm b}$ is the scalelength of the bulge and $K_0$ a modified
Bessel function. A cutoff to the bulge density is taken at
eight bulge scalelengths. For the dark halo the usual pseudo
isothermal sphere is assumed with a density ${\rho}_{\rm h}$ of
\begin{equation}
{\rho}_{\rm h} = {\rho}_0^{\rm h} \left[ 1 + \frac{R^2}
{R^2_{\rm core}} \right]^{-1},
\end{equation}
and belonging rotation velocity $v_{\rm h}$ of
\begin{equation}
v_{\rm h} = v^{\rm h}_{\rm max} \sqrt{ 1 - \frac{R_{\rm core}}{R}
{\rm arctan}\left( \frac{R}{R_{\rm core}} \right) },
\end{equation}
where $R_{\rm core}$ is the core radius of the halo related to
the maximum rotation $v^{\rm h}_{\rm max}$ by
\begin{equation}
v^{\rm h}_{\rm max} = \sqrt{ 4 \pi G {\rho}^{\rm h}_0
R^2_{\rm core} }.
\end{equation}

As template for the model galaxy the beautiful face-on spiral NGC 628
is used. It has a regular spiral structure and certainly not a bar.
This gives the possibility to compare the structure resulting from the
simulations with an actual galaxy in an unambiguous way. It is not
intended to make a model of NGC 628 in all its details, but only
to generate a galaxy which has roughly the same dimensions.
Photometry of NGC 628 is presented by Shostak \& van der Kruit (1984)
and by Natali et al. (1992). The first authors give a disc
scalelength for the Kodak J emulsion of 85\arcsec\ while the second
authors give $h_{\rm d} = $ 73\arcsec\ in the R-band. As a compromise
I shall use a value of 80\arcsec\ which gives for an adopted distance
of 11.5 Mpc a value of $h_{\rm d} =$ 4.5 kpc. A photometric disc
profile with that scalelength has been subtracted from the photometry
of Natali et al. and to the remaining central bulge light an exponential
has been fitted. From that follows a bulge scalelength of 7\farcs{12}
= 397 pc and the surface brightness of bulge and disc are equal
at a radius of {15}\farcs{3}. This photometric analysis gives the absolute
scales and brightness ratio of bulge and disc but not yet the masses
of these components.

NGC 628 is observed in \hi by Wevers et al. (1986) who give the radial
\hi gas density. This profile is supplemented with the H$_2$ radial
profile as observed by Combes \& Becquaert (1997) and the total
is multiplied with a factor 1.4 to account for Helium. The resulting
radial gas profile as it has actually been used is given in Fig. 2.
Note that although the addition of H$_2$ in the centre looks quite
impressive, it only contributes 8.5\% to the total gas mass
of 18.8~10$^9$ $M_{\sun}$.

%Fig3
\begin{figure}
\resizebox{\hsize}{!}{\includegraphics{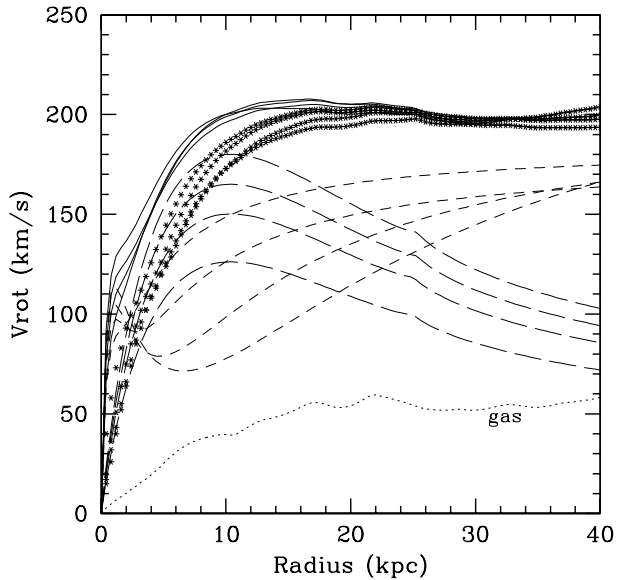}}
\caption[]{
For the photometry and gas content of NGC 628 a decomposition
has been made of an adopted rotation curve reaching a flat
level of 200 \kmss (full drawn line). Four different disc masses are
considered corresponding to maximum disc rotations of 
126, 150, 165, and 180 \kmss (long dashed lines). The combined
halo plus bulge rotation (short dashed lines) is then adjusted
to fit the adopted total rotation. For each disc contribution
the stellar rotation is given by the asterisks.
}
\end{figure}

Because NGC 628 is face-on its rotation curve cannot be
measured. Yet for its absolute luminosity and Hubble type the
Tully-Fisher relation of Rubin et al. (1985) gives a maximum
rotation of approximately 200 \kms. This value has been adopted as the
exact maximum rotation for the model galaxy together with the
observational establishment that rotation curves are flat in the
outer regions of a galaxy. For the exponential
disc rotation curves have been
calculated (Casertano 1983) scaled to four different maximum
rotational values of 126, 150, 165, and 180 \kms. These four
stellar discs all have a certain mass and mass-to-light ratio.
Combined with each disc is a bulge density distribution where
equal M/L ratios of bulge and disc have been assumed. 
For the four cases a dark halo has been created such that the total
rotation is approximately flat at 200 \kms.
The result of this fitting procedure is shown in Fig. 3 and
numerical values are presented in Table 1. Also given in Fig. 3 is
the actual stellar rotation which results after the subtraction
of the asymmetric drift, in this case for a $z_0$ value of 900 pc.
\section{Practical properties and adaptations}
\subsection{Main parameters}
To gain insight into the possible morphologies
of a disc, the main structural parameter to be investigated is the
ratio of maximum disc rotation to total rotation: $\vmd / \vmo$.
Though it has already been known since the calculations by
Ostriker \& Peebles (1973) that a galaxy where the disc approaches its
maximum disc limit is unstable to bar formation, (see also Bottema \&
Gerritsen 1997) that notion should be confirmed for an actual
3D disc with gas and star formation. Inevitably related to the matter
of stability must be the thickness of the disc. A thicker
disc has a larger vertical velocity dispersion.
The dispersions in the three directions are related, which
means that a thicker disc has a larger $Q$ value and should
at least be more stable to local distortions. Therefore $z_0$ is the
second main parameter to be investigated.

The precise amount and the process by which mechanical SN energy
is transferred to the gas is not known. Consequently
the influence of the parameter $\epssn$
will be considered in some detail. A fourth parameter has been added
during the course of the simulations. For a resulting SFR of 4
$M_{\sun} {\rm yr}^{-1}$ it appeared that the gas reservoir
becomes depleted pretty fast. To create a constant situation over
a reasonable time span, gas has been added at several rates.
Summarizing, the main parameters investigated in this paper are:
$\vmd / \vmo$, $z_0$, $\epssn$, and gas supply.

\subsection{Particle masses and numbers}
For most simulations the number of stellar particles was 150.000 and
the number of gas particles 16.000, though some simulations are presented
using 40.000 gas particles. In Table 1 the total stellar disc masses
are given for the various maximum disc rotations. From
$\vmd$ = 126 \kmss to 180 \kmss the stellar particle masses then
range from 0.3~10$^6$ to 0.6~10$^6$ $M_{\sun}$.
For a total gas mass of 18.8~10$^9$ $M_{\sun}$ and 16.000 gas particles,
the mass of one gas particle amounts to 1.18~10$^6$ $M_{\sun}$. During
the star formation process new stellar particles are created with masses
of $\frac{1}{2}$, $\frac{1}{4}$, and $\frac{1}{8}$ of the mass
of a gas particle, being comparable to the masses of the already
existent stellar particles. 

\subsection{The star formation rate}
The SFR as defined here is the total amount of gas mass which
is transformed into stellar mass, per time unit. In the code
no stellar mass is returned to the gas reservoir. For an actual
galaxy with a standard IMF, approximately 70\% of the gas
that has formed stars is eventually locked up in old stars
and stellar remnants. The other 30\% is returned, of which most
in a short time scale after SF. Because this time scale is so short
the non returning of gas in the numerical simulations is a good
approximation. However, the star formation rate that could be observed
is then not equal to the SFR as defined above. For a lock up rate
of 70\% the ``observable SFR'' is defined as 1/0.7 times the SFR
and can as such be compared with observations. Kennicutt (1983)
has determined an ``observable SFR'' of 3.8 $\msolperyr$
for NGC 628, when it is put at a
distance of 11.5 Mpc. There is a systematic
error of approximately a factor of two associated with this
value of 3.8 $\msolperyr$ mainly caused by the uncertainty of the IMF. 

\subsection{Stochastic heating}
An average galactic disc is essentially collisionless.
In a simulation a disc is made up of far less particles than the
number of stars in a real galaxy. As a result encounters between
particles will occur in simulations and the particle ensemble will
be heated up. This process of particle scattering is commonly
referred to as stochastic heating. In practice this means that a model
disc made up of a limited number of particles will heat up, get
thicker, and become more stable during a simulation.

In a real galaxy which contains gas, stellar heating may be generated
by two processes. At first heating by encounters of stars with
molecular clouds (Spitzer \& Schwarzschild 1953; Villumsen 1985).
Secondly a more effective mechanism seems to be heating
by transient spiral structures (Barbanis \& Woltjer 1967; Carlberg \&
Sellwood 1985) though in practice a combination of both mechanisms
is likely taking place. In a simulation with a limited number
of stars, and with spiral arms and cloud complexes both,
stochastic heating and real heating will take place. The problem
is to disentangle the two.

%Fig4
\begin{figure}
\resizebox{\hsize}{!}{\includegraphics{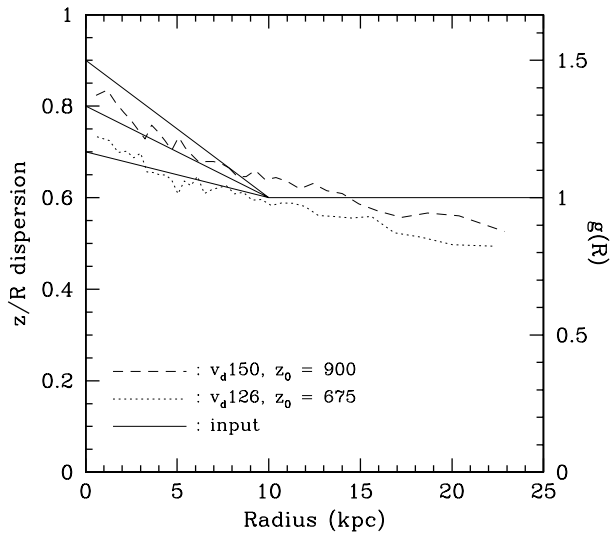}}
\caption[]{
The ratio of vertical to radial stellar velocity dispersion
of the model galaxy. The dynamical equilibrium is indicated
for two cases: $\vmd$ = 150 \kmss with $z_0$ = 900 pc and 
$\vmd$ = 126 \kmss with $z_0$ = 675 pc. Simulations are started
with the functionality given by the ``input'' line with
a $z/R$ dispersion at $R=0$ of 0.9, 0.8, 0.8, and 0.7 for
$\vmd$ = 180, 165, 150, and 126 \kmss respectively.
}
\end{figure}

This disentanglement can be done by considering the change of the
stellar velocity dispersion. From basic principles one can derive
for stochastic heating that
\begin{equation}
\frac{d \langle v^2 \rangle}{dt} \propto \frac{{\rm ln}N}{N},
\end{equation}
such that the change of the square of the velocity dispersion is
proportional to a function of the number of particles $N$. During
a simulation the square of the velocity dispersion will increase
by stochastic heating and by heating of clouds and arms.
If the same simulation is repeated for different numbers of stellar
particles both heating processes can be disentangled. Another way to
study the matter is by comparing the same simulations with and
without gas, as long as the gasless simulation produces a smooth
featureless disc. Using these methods, a small investigation was done
for a simulation using $\vmd$ = 150 \kmss and $z_0$ = 675 pc.
It appears that for a total of 180.000 stellar particles the disc
does not show stochastic  heating at a radius of one scalelength
and shows it barely at two scalelengths. As a final compromise
a number of 150.000 stellar particles has been used for all discs.
This means that beyond a radius of approximately two scalelengths a
small amount of heating will be present, not occurring in real galaxies.

%Fig5
\begin{figure*}
\resizebox{\hsize}{!}{\includegraphics{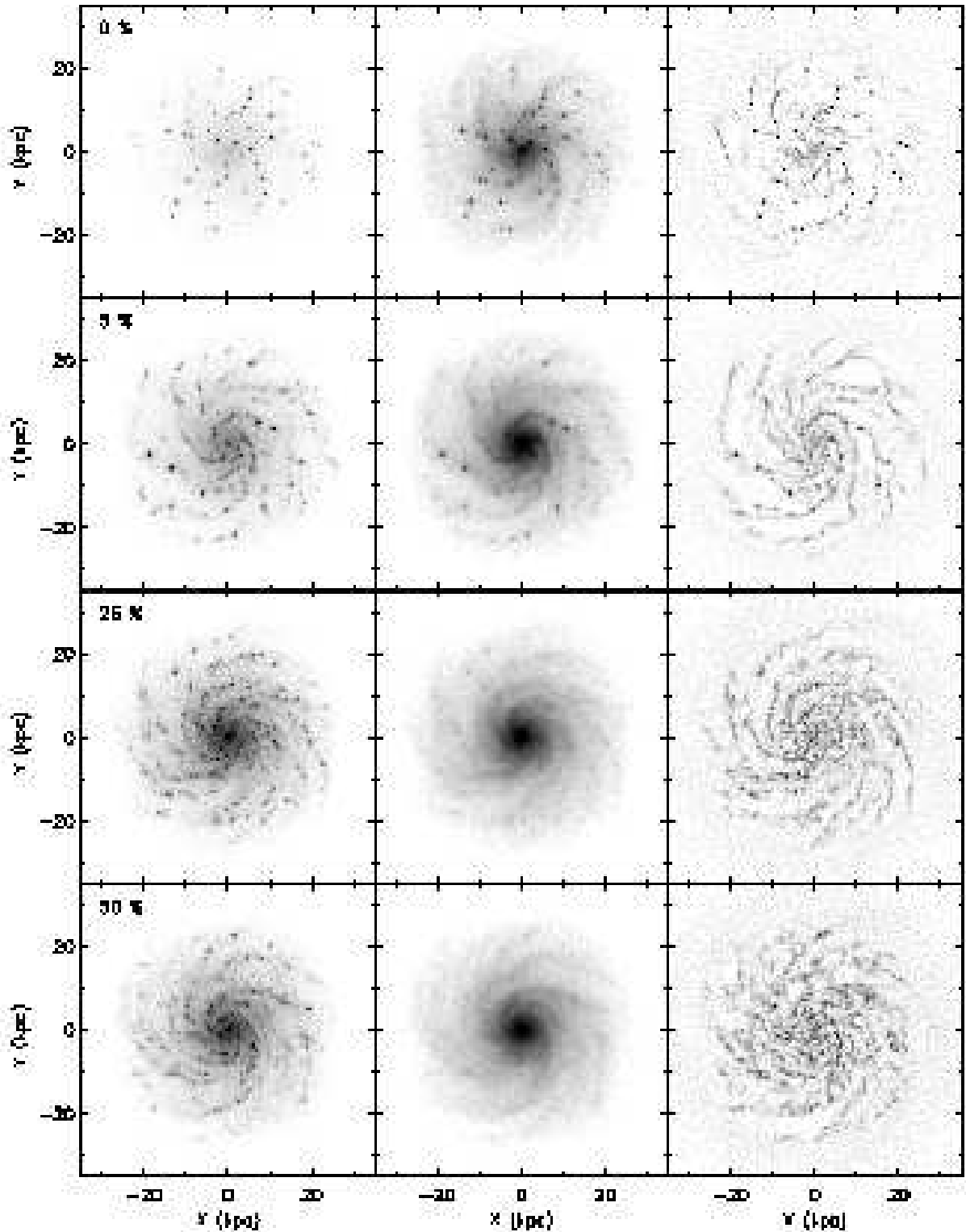}}
\caption[]{
The resulting galaxy after 1.5 Gyrs of simulation. In this and
following similar figures the left column gives the simulated
V-band image, the middle column gives the stellar mass image and
the right column the gas mass. The disc has a maximum rotation
of 126 \kmss and $z_0$ of 675 pc while gas was supplied at a rate
of 4 $\msolperyr$. From top to bottom row the mechanical SN feedback
efficiency was at 0\%, 5\%, 25\%, and 50\%. In case of none or low
SN feedback unstable giant SF regions appear. More SN feedback results
in a galaxy which is in SF equilibrium having an appearance not
unlike that of a real galaxy. High SN feedback destroys the coherent
gas spiral structure. 
}
\end{figure*}

\subsection{The softening length}
When performing 2D simulations of discs, the thickness
is generally artificially implemented by taking a stellar
softening of size comparable to the scaleheight of the disc.
The particles then do not meet in the plane but a disadvantage
of this adaptation is a considerable additional stabilization
compared to a 3D situation with the same dimension
and kinematics (Romeo 1994).

For 3D simulations considerations on the size of the 
softening length have been made by Hernquist (1987). It appears
that in order to achieve a good approximation of the force
computation the softening length should be smaller then
$\lambda$, being the mean interparticle distance at the half
mass radius. For an exponential disc Hernquist gives an approximation
of $\lambda \approx 2.63 ( N/h^2 z_0 )^{-1/3}$ which for the
present case of NGC 628 with $N = 150.000$ gives $\lambda \approx$
120 pc. One may consider other evaluations of $\lambda$, for instance
the mean interparticle distance at the half mass height 
above the plane. In that case $\lambda$ ranges from $\sim$ 110 pc
at the centre to $\sim$ 230 pc at $R = 2h$. When a bar develops, however,
$\lambda$ may become even smaller in the bar region. 
Therefore, for the galaxy under investigation in this study the 
softening length should be smaller than approximately 100 pc. 

To get a lower limit on the softening length is more problematic. 
If $\varepsilon$ is too small, strong interactions between particles
will occur. That is especially problematic when particles with
different mass are present; the lightest particle will then
be ejected, carrying with it large amounts of energy and angular
momentum. In addition, for close particle interactions the time 
step may be too small to achieve a correct orbit integration. 
To investigate matters, pure stellar simulations have been performed
for a case where a moderate bar develops (Fig. 10, bottom left) with
$\varepsilon$ ranging from 2 to 200 pc. For $\varepsilon$ between
10 and 200 pc the resulting structure and stellar kinematics is
practically equal. For $\varepsilon \la 10$ pc particles get ejected
from the galaxy (particles have masses differing $\sim$ a factor two)
which changes the eventual morphology. It is therefore concluded
that for the present calculations there is a safe range between
10 to 100 pc for the stellar softening length. It has been fixed
at 20 pc for all simulations. In no case there will then be artificial
stabilization by a too large softening length. 

\subsection{The quotient of vertical to radial stellar velocity dispersions}
In first instance this quotient was put at 0.6 for all radii as an
extrapolation of the number for old disc stars in the solar
neighbourhood. Consequently the function $g(R)$ in Eq. (12)
$\equiv 1.0$ for all radii. During the initial stages of
the simulation, for the more massive discs, there appeared
to be some outflow of stars from the central regions. A solution
to this problem was found after comparing the final quotient
for some realistic simulations with the original input (see Fig. 4).
At the end of the simulations the velocity dispersion ellipsoid is
always more spherical in the inner regions, while it remains at
$\sim 0.6$ near two scalelengths. The initial model galaxy has been
adapted to have a function $g(R)$ slightly variable as a function
of radius. As shown in Fig. 4, the central quotient was put at
0.7 (for $\vmd = 126$ \kms), at
0.8 (for $\vmd = 150$ and 165 \kms) or 0.9 (for $\vmd = 180$ \kms)
and decreased linearly to 0.6 at a radius of 10 kpc. For such a 
setup the initial central instability has completely disappeared.
For a disc with maximum disc rotation of 126 \kmss or lower this
adaptation was, in principle, not necessary; the galaxy slowly changes to 
the final quotient without noticeable redistribution of matter.

\section{Creating a galactic disc in equilibrium}
\subsection{The amount of SN action}
The first parameter of which the effect has been investigated is
the SN efficiency. To that aim a realization of the model galaxy
has been chosen which is close to what observations of galactic discs
suggest. Therefore a thickness $z_0$ of 675 pc is assumed
(van der Kruit \& Searle 1982) and a disc maximum rotational 
contribution of 63\% (Bottema 1993). By no means the statement is given
that this galactic constitution is the closest to reality; only
that it is a reasonable assumption and suitable to test the effect
of the amount of SN feedback. During the simulations gas has been
supplied at a rate of 4 $M_{\sun} {\rm yr}^{-1}$ being approximately
equal to the amount of gas depletion by SF.

Results of $\epssn$ = 0\%, 5\%, 25\%, and 50\% are given in Fig. 5 where
for each case the simulated V-band image, stellar mass image, and
gas mass image is presented. In fact, Fig. 5 is pretty self explainable.
If the SN feedback is absent or small, dense clumps of gas and young
stars are created, which dominate the V-band and gas image. On these
clumps cold gas continues to accumulate and because the density is
so large, the cooling outruns the heating by the FUV field of young
stars. It should be kept in mind that cooling proceeds proportional
to the square of the gas density (Eq. 4) and heating proportional to the 
first power of the density (Eq. 6) and a runaway process as observed
might indeed be expected to occur. For an $\epssn$ of 25\% the clumping
process has almost completely disappeared and a number of spiral arms
appear in the stellar light, mass, and gas distribution. Certainly
in the gas structure these arms have a certain width contrary to 
the lower $\epssn$ simulations where the gas arms are thin and have a
threaded appearance. Increasing the supernova feedback to the high value
of 50\% results in the creation of multiple holes in the gas structure.
The feedback is now so violent that spiral arms in the gas image have
all but disappeared. As a result also the strength of the spiral
arms in the light and stellar mass images decreases.

%Fig6
\begin{figure}
\resizebox{\hsize}{!}{\includegraphics{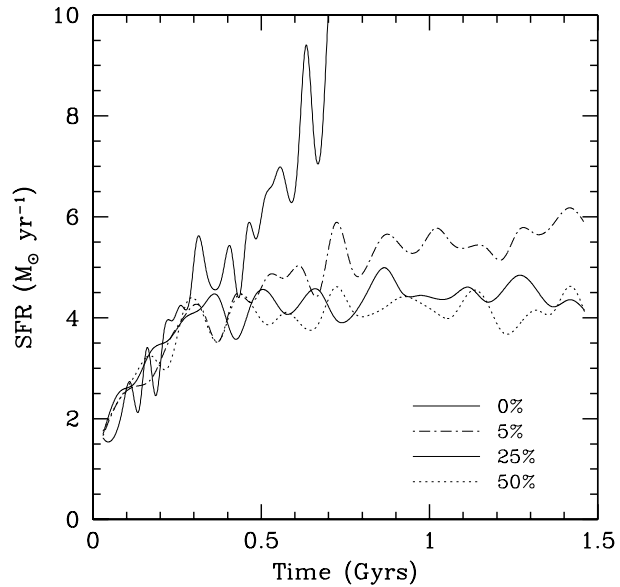}}
\caption[]{
The star formation rate during the simulations presented in Fig. 5.
For low efficiencies of the SN feedback (0 and 5\%) SF is boosted
in numerous giant SF regions. A larger SN feedback results in an
equilibrium situation at which the SFR reaches a constant level
determined by the gas supply and cooling function.
}
\end{figure}

For the different SN feedback situations the resulting SFR as a function
of time is presented in Fig. 6. In case of low $\epssn$ and the 
generated severe clumping the SFR keeps on increasing, evidencing the
runaway star formation process. If this would continue, the galaxy
gets quickly depleted of its gas and the SF activity would subside
again. A larger SN feedback results in a lower
SFR, though above $\epssn$ = 25\% the SFR remains more or less constant
at a level of 4.2 $M_{\sun} {\rm yr}^{-1}$. Corrected for stellar mass
loss (Sect. 4.3) this gives 6 $M_{\sun} {\rm yr}^{-1}$ which is in good
agreement with the observed value for NGC 628 (Kennicutt 1983).
There is no evidence for the process of stochastic self
propagating SF by supernovae explosions (Mueller \& Arnett 1976;
Elmegreen 1992).
On the contrary, more SN action decreases the amount of SF when 
the SN action is low, while for higher regimes of SN action the SFR
seems to be almost independent of the amount of SN feedback. During the
simulations star formation occurs mainly by cooling and dissipation
of the gas. SN action prevents a runaway SF process and influences
the gas morphology. 

An investigation has been made of how the structures and relations
of Fig. 5 depend on other parameters. It does not depend on the value
of $M_{\rm crit}$ (Eq.~7). There is a dependence, however, on the value of the
collapse factor $f_c$ (Eq.~8). If the time span before star formation
is increased from the nominal value of $0.5 \times t_{ff}$ the relative
amount of cold ($T < $ 500~K) gas increases and the clumping 
process sets in at higher SN feedback values, and vice versa. Since
the ratio of cold to warm \hi gas is as yet undetermined it is impossible
to pinpoint down the exact value of $\epssn$ (see Sect. 7 for an ample
discussion). Still, following indications of the observations leading
to the adopted values for the parameters, the relation between the
morphology in Fig. 5 and the value of $\epssn$ is likely accurate to
within a factor of two. 

%Fig7
\begin{figure}
\resizebox{\hsize}{!}{\includegraphics{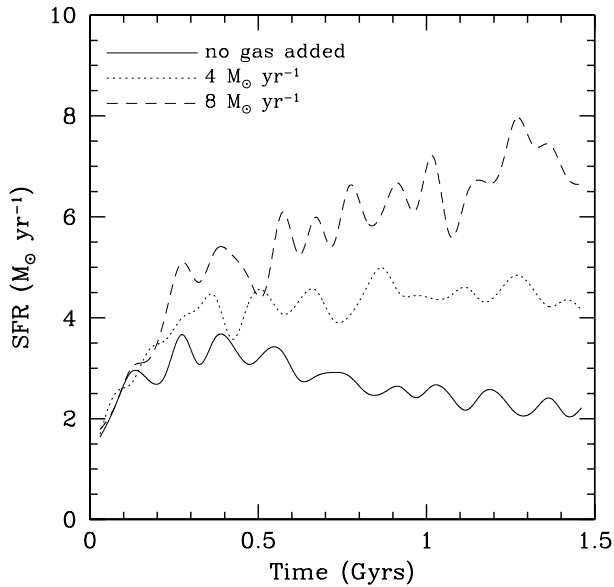}}
\caption[]{
The star formation rate during simulations with different amounts
of gas supply.
Eventually the SFR will adjust itself to the amount of gas
that is being added.
}
\end{figure}

\subsection{Different amounts of gas supply}
As noted before, it became evident during this investigation
that when having a gas consumption rate of $\sim$ 4 $\msolperyr$ 
for a total gas amount of 18.8~10$^9$ $M_{\sun}$,
the gas gets depleted pretty fast. In the mean time, it takes
a while, lets say some 0.8~10$^9$ yrs, before a stable galaxy
emerges (see Fig. 6). To have a constant gas reservoir over at least
this adjustment time it was decided to create the possibility
to add gas at a certain rate. For a real galaxy it is also expected
that gas gets accreted, although the way how is (still) unclear.
Therefore this gas suppletion is less artificial than it
might seem.

Care has been taken that the addition of gas does not alter the
available gas structure. For example, gas should not be supplied at
those positions where it has just recently been removed by star
formation. Gas particles have been added randomly by selecting another
gas particle and placing the new gas close to that selected particle.
From this selection are excluded gas particles with temperatures above
20.000~K or below 5000~K. Also excluded are, of course, SN particles
or particles already Jeans unstable. It became clear that when gas 
is added at places beyond
the stellar disc, that only increases the amount of gas over there, taking
more computing time, and not replenishing the gas where it is depleted.
Therefore no gas has been added at radii larger than 25 kpc. It was
originally expected that extra gas at large radii would
slowly migrate inwards by dissipation and viscous interaction.
That process, however, did not take place.

The morphology and SFR of a galactic disc 
has been investigated in cases of gas addition of 0, 4, and
8 $\msolperyr$ during a period of 1.5 10$^9$ years. 
A disc is chosen equal to the one where
the amount of SN action was investigated, with a fixed $\epssn$ of
25\%. A SFR at a constant level is achieved when 
$\sim 4$ $\msolperyr$ is added. When no gas is added the SFR approaches
4 $\msolperyr$ after 3 to 4~10$^8$ years from the start of the simulation,
and declines gradually thereafter (see Fig.~7). At the end of the 
simulation the SFR has decreased to 2 $\msolperyr$. As can be seen in
Fig. 7, when gas is added at 8 $\msolperyr$ the SFR keeps on increasing
to reach eventually the level of gas input. In this way one can
draw an important though not unexpected conclusion: the SFR in a disc
adjusts itself to the amount of gas that is being supplied. The
time scale on which this happens is for the present simulation 
$\sim$ 1 Gyr, but will likely depend on the numerical resolution.
When the images of the galaxy are inspected at 1.5 Gyrs
after the start of the simulation it appears that the morphology is only
slightly dependent on the amount of gas supply and consequently also
only slightly dependent on the actual SFR. More gas and higher SFR
results in a somewhat stronger and less regular spiral pattern. 
When 8 $\msolperyr$ is added also a few dense gas clouds develop.

\subsection{Comparison with observations}
The V-band images as presented in the left column of Fig.~5
are made to resemble a galaxy as depicted in an optical photograph.
To that aim the intensity is given on a square-root gray scale and cut off
(saturated) above a certain level. There is always some subjectiveness
involved in selecting the scale and cutoff and for the best result I
aimed at achieving the best overall contrast. In reality, however,
this will be equal to the process of developing and printing of
actual photographs.
A shortcoming of the simulations is the
lack of dust. Certainly the young SF regions in real galaxies are
often (at least partly) enshrouded in some amount of dust, a situation
that cannot be reproduced. Another problem is that every new star
particle (= SF region) has an IMF going up to stars of 125 $M_{\sun}$ 
which will completely dominate the light in every passband. In a real
galaxy there are only a few SF regions which contain such massive
stars. To compensate, at least in part, for the lack of dust and for
the discrete appearance of very massive stars, the V-band image is made
for all the stars at an age 3~10$^7$ years older than the actual
age of the stars. The net effect of this procedure is an image
less dominated by young star clusters. Admittedly, there is some
artificiality in creating a simulated V-band image to compare with
photographs, but there is no way around this.

When there is no SN action the V-band image is completely dominated
by bright \hii regions and to my knowledge such a galaxy does not
exist. For $\epssn$ = 5\% the optical image is also very knotty and
is barely consistent with the image of NGC 628 on, for example,
page 99 of the Shapley-Ames catalog. When $\epssn$ is larger the optical
images are not unlike images of real galaxies. The problem with the stellar
mass images is that these cannot be observed. Even in the near
infrared passbands young populations will supply extra light, e.g. have
a smaller M/L ratio compared to an old population.

Contrary to the optical and stellar mass image, the image of the gas
can actually be compared with observed \hi structures. High resolution
\hi observations of nearby late type galaxies are presented by a.o.
Braun (1995) and Kamphuis (1993). 
Such a high resolution is necessary because one would like to compare
the small scale gas structure and in particular the size and distribution
of holes created by SN feedback. For example, Fig. 5 lower right
panel shows lots of small holes and cavities in the inner region, all
created by SN explosions. In addition, a particular feature of SN
feedback are holes blown into the ridge of spiral arms. This can
be witnessed in the same image at for example positions (x,y) =
(5,-15) and (-4,19). The observations of the gas structure do never
show the knotty and threaded gas structures as for $\epssn$ = 0
and 5\%. The best resemblance between simulations and observations
is for $\epssn$ around 25\%, though also more amorphous gas structures
resembling the structure for $\epssn$ = 50\% are observed. A complicating
factor is that some of the nicest \hi spiral structures are observed
in galaxies with grand design spiral arms obviously caused by an interaction
with other galaxies, for example M81 (= NGC 3037). In such cases gas
becomes aligned with the gravitational spiral disturbance, influencing
its global appearance. This should be taken into account when
comparing the simulated gas structures with real observations.

\subsection{Whence spiral arms?}
The spiral structure which is generated in the present 
simulations of an isolated galaxy is the result of an equilibrium
of four actions. On the one side cooling and dissipation accumulates
the gas in threads and clumps. This is counteracted by FUV heating 
and SN feedback. If the latter action is too small the gas collapses
completely into a few giant SF regions. On the other hand, if the action
of heating and SNs is too large, warm gas get dispersed over the disc
preventing the formation of any coherent structure.
Except for this amount of mechanical feedback the resulting
gas and mass structures do not depend on details (like IMF,
$M_{\rm crit}$, etc) of the star formation process (see Sect. 5.1
and GI97). It is therefore likely that these structures do not
depend on the exact star formation recipe. 

The behaviour of the gas is the main cause of
structure formation. As a result the underlying stellar mass
distribution becomes disturbed and also in the stellar disc spiral
arms are formed. In this way gas mass enhancements more or less
swing amplify the stellar mass into a spiral structure. But this
is probably not the whole story. If it were only the need of mass
accumulations to generate spirals, then it is expected that the
low $\epssn$ situation with its specific clumps would generate
the strongest spirals. Obviously that is not the case. The gas 
enhancements really have to be wound up in spirals by the differential
rotation to create an accompanying mass enhancement. A situation
is established where shearing gas filaments swing amplify
the underlying mass distribution into a transient small
scale density wave. The latter appears as a spiral arm.
More insight is gained when the 
simulation is played as a movie on the TV screen. It can then be
noticed that coherent spiral features develop and exist for
approximately half a rotation period. Spiral arms are ``wound''
into existence and disappear by winding and stretching too much
or by interaction with other spiral features. Such interactions also
create the typical bifurcation patterns of the arms often
seen in real galaxies. Not created in the present simulations
are the grand design m=2 spirals witnessed in a few exemplary galaxies.
These matters will be further investigated and discussed in 
sections 8 and 9.

%Fig8
\begin{figure*}
\resizebox{\hsize}{!}{\includegraphics{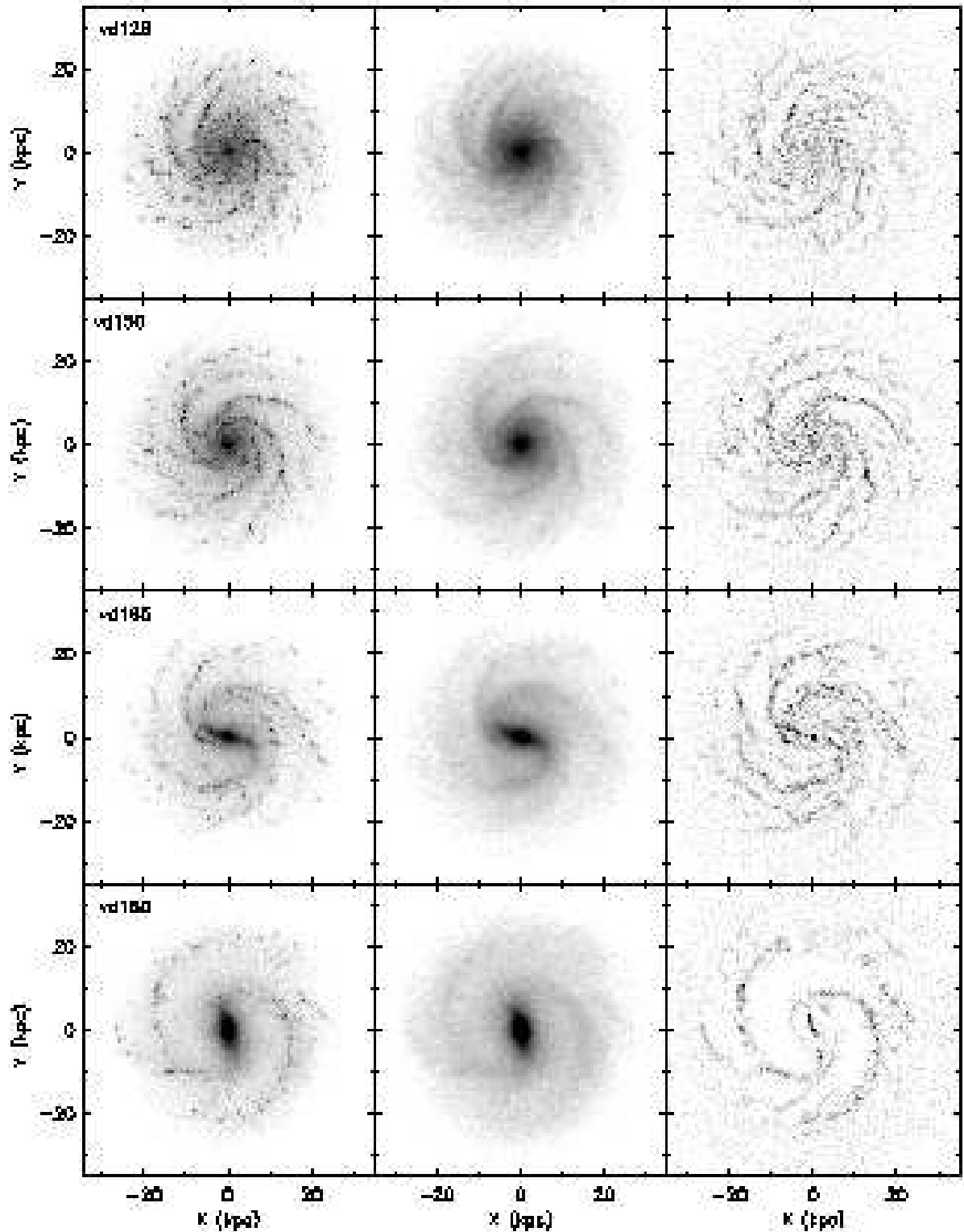}}
\caption[]{
For a fixed SN efficiency of 25\%, gas suppletion rate of
4 $\msolperyr$, and $z_0$ = 900 pc, the resulting galaxy after 1.5 Gyrs
of simulation is presented for four values of the maximum
rotation of the disc (126, 150, 165, and 180 \kms). Less massive
discs evolve into a regular spiral galaxy, while more massive
discs develop a bar. In case of a bar, gas is transported inwards as
can be noticed in the right column.
}
\end{figure*}

\section{The bar instability}
\subsection{The appearance as a function of disc mass}
An amount of SN feedback has now been determined which results in 
a good resemblance with observed structures in a disc while for a gas
addition of 4 $\msolperyr$ the gas reservoir remains constant. 
Fixing these parameters and fixing the thickness parameter $z_0$
at 900 pc the resulting morphology for different disc contributions
to the total rotation has been investigated. Images of the galaxy
after 1.5~10$^9$ years of simulation are presented in Fig. 8 for
$\vmd$ of 126, 150, 165, and 180 \kms. As for Fig. 5, Fig. 8 speaks
pretty much for itself. Discs with a moderate rotational contribution 
of 126 and 150 \kmss evolve into a regular spiral galaxy. For
$\vmd$ = 150 \kmss there appear to be less arms which are more pronounced
than in the case of $\vmd$ = 126 \kms. However, the images presented
in Fig. 8 are only one snapshot at one particular time. During the
simulations arms come and go and are variable in strength. Judging
the whole simulation from $t=0$ to $t =$ 1.5~10$^9$ years the 
$\vmd$ = 150 \kmss case has also on average a slightly more pronounced
spiral structure. 

Discs with a larger rotational contribution obviously develop a bar. 
For $\vmd$ = 165 \kmss the disc starts out by forming a few large
and broad spiral structures, a situation which continues to
$\sim$ 1.2~10$^9$ year. Then quite suddenly the bar appears. 
For $\vmd$ = 180 \kmss the simulation also sets out by forming
broad dominant spiral features which already after $\sim$ 0.5~10$^9$
years settle into a barred structure. When the bar is present gas is 
transported inwards rapidly along the bar. The same conclusion was
reached by Athanassoula (1992) for simulations of gas moving in a 
rigid barred potential. That study shows that the gas is transported
inwards along shocks appearing on the leading side of the bar. 
The present calculations give the same results, though the resolution
in not good enough to investigate details of this process.

%Fig9
\setlength{\unitlength}{1cm}
\begin{figure*}
\begin{minipage}{11.4cm}
\begin{picture}(11.4,10.9)
\resizebox{11.4cm}{!}{\includegraphics{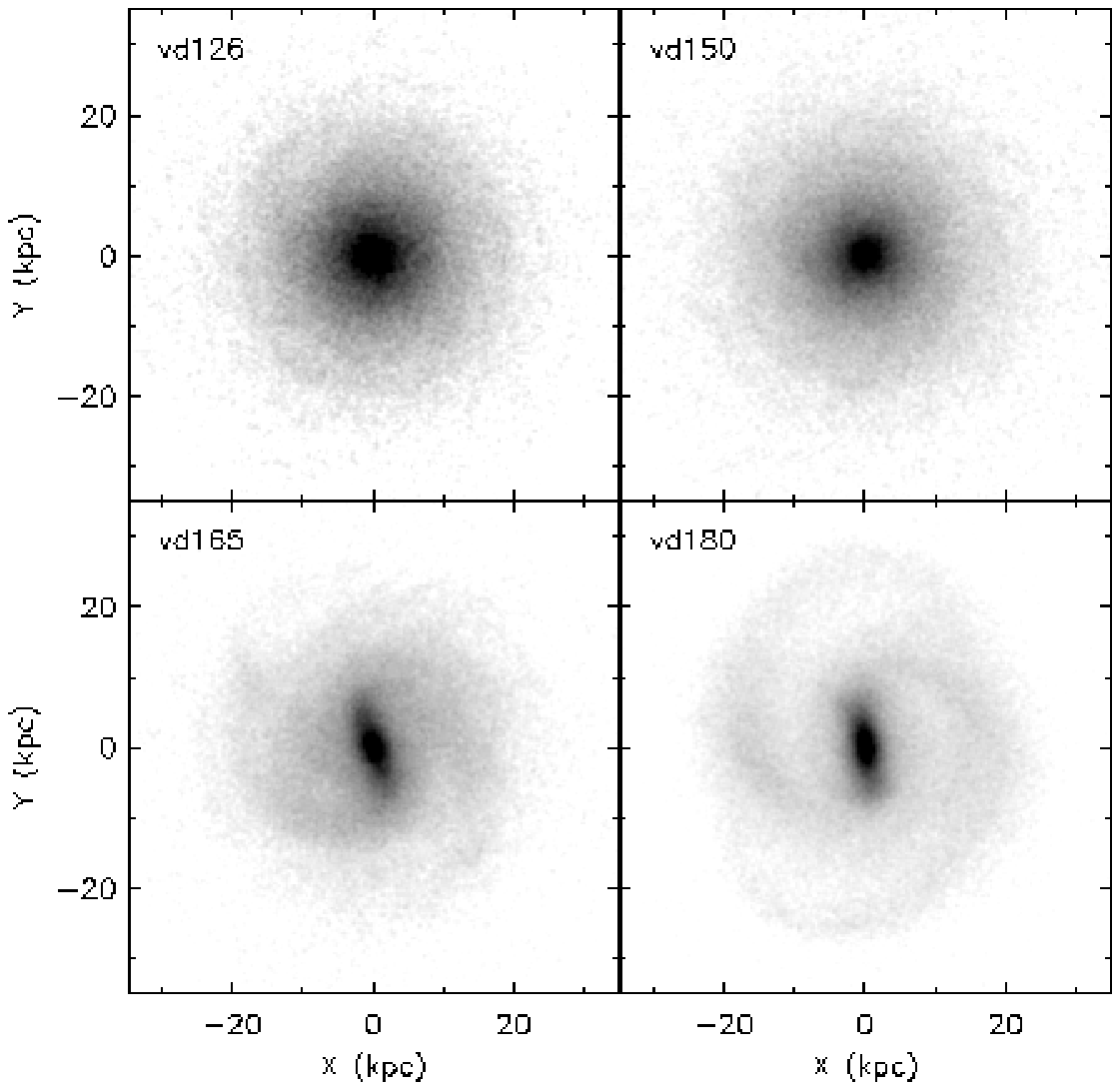}}
\end{picture}
\end{minipage}
\hfill
\begin{minipage}{5.7cm}
\begin{picture}(5.7,5.1)
\end{picture}
\caption[]{
The stellar mass image of the same four galaxies presented
in Fig. 8 after 1.5 Gyrs of simulation, but now the gas is replaced
by stars. Without gas the bar does, or does not, appear just as well.
It is obvious, though, that gas is needed to generate spiral structure. 
}
\end{minipage}
\end{figure*}

In the bar the gas density is large, cooling is fast and new stellar
particles form in appreciable quantities. So much young stars form
at some positions and certainly at the centre that FUV heating 
outweighs the cooling and gas remains at $\sim$ 10$^4$~K while
becoming ever more dense.
When gas particles accumulate within the softening length 
numerical problems occur. To avoid these, the gas was forced
to form new stars above a density limit of 10 \msol\ pc$^{-3}$
even though the star formation criteria of subsection 2.4 were
not yet satisfied. It was not intended and it is not possible
to simulate the processes going on at the very centre of the 
bar region. What the simulations do show is that gas is
transported inwards along the bar, where it forms stars at a 
considerable rate. As a consequence a concentration of young stars
emerges at the inner few kpc which takes on a more or less spherical
shape. This can be called the development of a bulge. For the whole
galaxy the radial luminosity profile changes from exponential to
Freeman type II (Freeman 1970) where the shoulders of the profile
coincide with the bar.
Getting back to Fig. 8, one can notice
the effect of the inwards gas transport. For $\vmd$ = 180 \kmss
the bar has existed longer than for $\vmd$ = 165 \kms.
As a consequence in case of $\vmd$ = 180 \kmss the region
around the bar has already been severely depleted of gas.
The results of the present simulations of the barred galaxies
are identical to those of Friedli \& Benz (1995). They also
specifically report the gas inflow, SF along the bar and in the
centre, and the formation of a bulge.

But, do we need the gas to form or not to form bars? To investigate
this, the simulations presented in Fig. 8 have been re-run
without gas. In order to obtain nearly the same density structure and
same rotation curves the gas has been replaced by stellar particles
which were given the same sech$^2$ vertical distribution as
the other stars. The resulting density structure after 1.5~10$^9$ years
is given in Fig. 9. For $\vmd$ = 165 and 180 \kmss the structure
closely resembles the mass structure in case  gas in included. Only
the central regions are less pronounced because no (gas) mass has been
transported inwards. Anyway, the conclusion is straightforward; gas is
not essential for the formation of bars, at least for the galactic
constitution investigated presently. Yet, when inspecting Fig. 9 
for the cases of $\vmd$ = 126 and 150 \kmss another fact is obvious. 
Namely that gas is essential for the formation of spiral arms. 
If there is no gas the galactic disc is featureless, while if gas is
included, the top two rows of Fig. 8 show that a rich spiral
structure may develop. 

\subsection{$Q$}
Although in principle not applicable for global instabilities,
Toomre's (1964) $Q$ criterion is often invoked to assess the
dynamical state of a galactic disc. When rigorously applied,
$Q$ can then serve as a kind of galactic thermometer, but this
rigorous application has its problems as we shall see. Toomre's $Q$
value indicates when an infinitely thin pure stellar disc becomes
unstable to local instabilities. It is defined as
\begin{equation}
Q = \frac{\dispr \kappa}{3.36 G \sigma},
\end{equation}
where $\sigma$ is the stellar surface density and $\kappa$ the 
epicyclic frequency given by
\begin{equation}
{\kappa}^2 = \frac{2 v_c}{R} \frac{\partial v_c}{\partial R}
+ \frac{2 v_c^2}{R^2},
\end{equation}
where $v_c$ is the circular test particle rotation. $Q$ is thus
defined there where the epicyclic approximation holds, meaning
$\dispr < v_c$. This leads to the first reason why the $Q$ criterion
does not apply for bars. Bars develop in the central regions of
galaxies which have a large disc mass contribution. At exactly these regions
and for those circumstances the epicyclic approximation clearly
does not hold. The second reason is, of course, that it is a local
and not a global criterion. 

To give the $Q$ thermometer
for the four simulations of Fig. 8 it has been calculated as a function
of radius. This calculation involves a third problem; what to use
for $v_c$ and for $\sigma$ in case of a disc with both stars and gas?
If dispersions are small the stellar rotation $(v_{\ast})$ is
nearly equal to the testparticle ($\sim$ gas) rotation 
and it does not matter whether one inserts $v_{\ast}$ or 
$v_{\rm gas}$ for $v_c$.
But in the inner regions of massive discs the asymmetric
drift is large and it does matter for the value of $Q$ which rotation
is used. Inclusion of a gas component will decrease the effective
$Q$ value as demonstrated for a thin disc by Bertin \& Romeo (1988).
For a real galaxy a straightforward recipe for the conversion of
$Q$ so as to include the gas cannot be given. To make a first
attempt one might substitute ${\sigma}_{\ast} + {\sigma}_{\rm gas}$
for the surface density $\sigma$ in Eq. (20). In this way I was lead
to define two $Q$ values, a pure stellar one ($Q_{\ast}$) defined as
\begin{equation}
Q_{\ast} = \frac{\dispr {\kappa}_{\ast}}{3.36\; G {\sigma}_{\ast} },
\end{equation}
and a $Q_t$ given by
\begin{equation}
Q_t = \frac{\dispr {\kappa}_{\rm gas}}{3.36\; G ({\sigma}_{\ast}
+ {\sigma}_{\rm gas}) },
\end{equation}
the two of
which should give an unambiguous description of the value and range
of the $Q$ thermometer. The result is given in Fig. 10.
The replacement of $v_{\ast}$ by $v_{\rm gas}$ has an effect on
$Q$ only at the inner regions. On the other hand, for large radii
and especially for the least massive discs the inclusion of the
gas surface density leads to a substantial decrease of $Q$.
Only the two most massive discs form a bar. Reading the $Q$ thermometer
in Fig. 10 it can be noticed that for these galaxies both
$Q_{\ast}$ and $Q_t$ fall below 1.2 at certain inner radii. It is
also at these positions where the bar eventually develops.
Further simulations with different galaxy constitutions (next
subsection) show that a level of $1.2 \pm \sim 0.1$ 
for any minimum $Q$ is indeed
critical for bar (in)stability.

There is a small complication, however. As noted in subsection 4.6,
an initially constant vertical to radial velocity dispersion ratio of 0.6
changes into a different functionality expressed in the function
$g(R)$ of Eq. (12) and displayed in Fig. 4. For discs with
$\vmd \gid$ 150 \kmss the setup has been adapted to accommodate
this changing dispersion ratio. For $\vmd$ = 126 \kmss this was
not necessary because that situation settles gently, without mass
transport into a stable state with a specific $g(R)$. But a different
dispersion ratio results in a different $Q$ value. If the disc mass
is fixed, $Q$ can be expressed as
\begin{equation}
Q \propto \frac{\sqrt{z_0}}{g(R)},
\end{equation}
which means that if $g(R)$ is larger than 1.0 as it is in the inner
regions, $Q$ is smaller than when the dispersion ratio were 0.6 
everywhere. As a consequence, for $\vmd$ = 126 \kmss $Q$ evolves
to slightly lower values a demonstrated in Fig. 10 by also showing
$Q(R)$ after 1.5 Gyrs. For $\vmd$ = 150 \kmss the radial functionality
of $Q_{\ast}$ after 1.5 Gyrs is indistinguishable from the $t=0$ situation
demonstrating that the input $g(R)$ was close to its eventual value.
Equation (24) shows another
important relation. To increase $Q$ by making the disc thicker is not
easy. For example, to bring up $Q$ from 1.0 to a stable $Q=2$, the
value of $z_0$ has to be increased by a factor four. This changes
$h/z_0 = 5$ into $h/z_0 = 1.25$ and the disc is in principle
not a disc any more.  

%Fig10
\begin{figure}
\resizebox{\hsize}{!}{\includegraphics{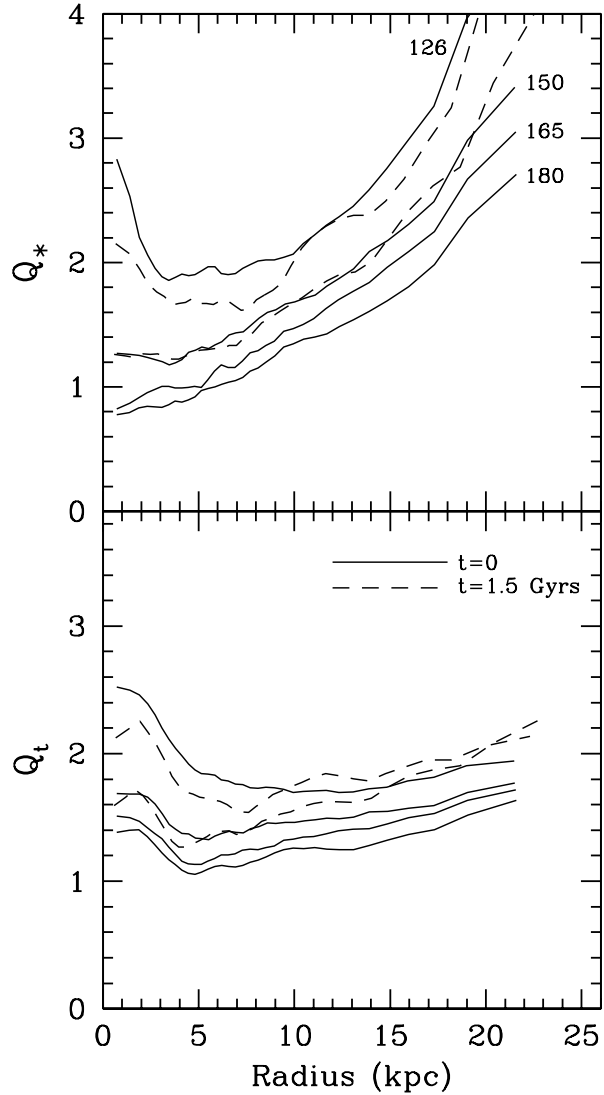}}
\caption[]{
The initial $(t=0)$ value of 
the $Q$ parameters ($Q_{\ast}$: Eq.~22 and $Q_t$: Eq.~23) for the four
galaxies of Fig.~8 ($z_0$ = 900 pc, $\epssn$ = 25\%). For the
stable discs $Q$ is also given at $t$ = 1.5 Gyrs by the dashed lines.
When any $Q$ falls below $\sim 1.2$ at certain radii a bar develops. 
}
\end{figure}

\subsection{A bar stability criterion}
The matter of bar (in)stability has been further investigated by
performing a number of simulations for discs with various masses
and thicknesses. As noted above, the stability depends on the 
precise value of $g(R)$ and as a consequence an initial situation 
should be close to having its equilibrium vertical to radial 
dispersion ratio. For the four different disc masses with belonging
maximum rotation a good approximation to this equilibrium
was found by the functionality given in Fig. 4.
In each case the thickness parameter $z_0$
was changed until stability changed into instability or vice versa.
%
%TABLE 2
\begin{table}
\caption[]{Bar stability investigation}
\begin{flushleft}
\begin{tabular}{lllll}
\hline
\noalign{\smallskip}
Disc & $z_0$ & $z_0$ & Bar & Time of bar \\
$v_{\rm max}$ & $t=0$ & $t=1.5$ Gyrs & length & appearance \\
(\kms) & (pc) & (pc) & (kpc) & (Gyrs) \\
\noalign{\smallskip}
\hline
\noalign{\smallskip}
126 & 300  & 450  & 5      & 0.6 \\
126 & 400  & 460  & 3.5    & 0.75  \\
126 & 500  & 530  & stable &  \\
126 & 675  & 675  & stable &  \\
126 & 900  & 900  & stable &  \\
150 & 500  & 790  & 8.5    & 0.75 \\
150 & 675  & 725  & 7      & 0.6  \\
150 & 900  & 940  & stable &  \\
165 & 900  & 1025 & 13     & 1.2 \\
165 & 1125 & 1280 & stable &  \\
180 & 900  & 1365 & 15     & 0.45 \\
180 & 1125 & 1540 & 16     & 0.75 \\
180 & 1300 & 1580 & 16     & 1.2 \\
\noalign{\smallskip}
\hline
\end{tabular}
\end{flushleft}
\end{table}

%new paragraph
The results of this investigation are summarized in Table 2 and
graphically in the $({\vmd},z_0)$ plane in Fig. 11. Even for
$\vmd$ = 126 \kmss a bar can be generated when the disc is made
thin and cool enough. On the other hand, for the maximum disc
case of $\vmd$ = 180 \kmss it seems impossible to avoid a bar.
For a very thick disc with $z_0$ = 1300 pc a bar still appears.
To make the disc even thicker, one needs large stellar velocity
dispersions and the epicyclic approximation cannot be applied
over most of the radial extent. A determination of the proper
asymmetric drift then becomes problematic. For the more massive
discs a long bar develops, while the size of the bar decreases
towards the less massive discs. The dependence of this bar length
on the thickness of the disc seems to be small. In this way one might
determine the disc contribution to the total rotation for a barred
galaxy by relating the length of the bar to the scalelength of the
disc. Unfortunately, the radial photometric profile of barred
galaxies is often of an extreme Freeman type II (Freeman 1970) making
it difficult to assign a scalelength. The simulations of barred
galaxies also show such an evolution towards type II and suggest
that the original scalelength can best be retrieved by taking
a global average of the eventual profile in the logarithmic domain.

%Fig11
\begin{figure}
\resizebox{\hsize}{!}{\includegraphics{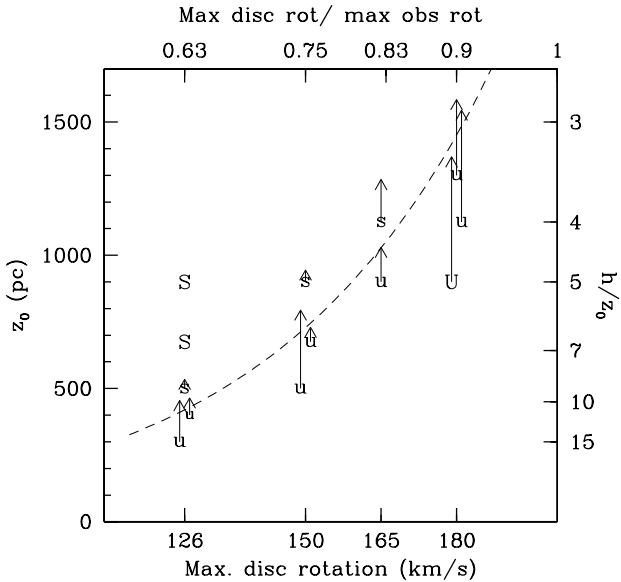}}
\caption[]{
Results of the bar stability investigation for discs with
various masses and thicknesses. A ``U'' means that a bar develops,
in case of an ``S'' a bar does not develop. The arrows indicate the
heating and following thickening of the stellar disc after 1.5 Gyrs.
A division of the $({\vmd},z_0)$ plane in a stable and unstable
section can be made according to Eq. (25) which is represented
by the dashed line.
}
\end{figure}

In Fig. 11 a number of sampling points are given alongside the 
instability threshold. Of course, more intermediate cases could
have been simulated but the results now gathered already give
a good impression. When the unstable (U) and
stable (S) data points of Fig. 11 are plotted in a log$(z_0/h)$ versus
${\vmd}/{\vmo}$ plane, a division by a straight line into a stable
and unstable section can be made. This dimensionless division line
is parameterized by
\begin{equation}
\frac{z_0}{h} = 5.17\; 10^{-3}\; {\rm exp}(4.592\; {{\vmd}/{\vmo}}),
\end{equation}
which has been converted to $({\vmd},z_0)$ and is plotted 
as the dashed line in Fig. 11. Extrapolation to $\vmd$ = 180 \kmss
indicates that a stable disc for that case should be achieved
approximately near $h/z_0 = 3.1$ or $z_0$ = 1450 pc, which represents
a puffed up disc rarely seen in reality (van der Kruit \& Searle 1982;
Kregel et al. 2002).

Obviously, this criterion is derived for one specific galactic 
constitution and its general, dimensionless applicability should
be further assessed. Does it hold, for example, for a different
$\vmo$? An indication that it indeed holds comes from the study
of Bottema \& Gerritsen (1997) where a similar stability change
over has been established for a disc with $\vmo$ = 120 \kms.
The applicability of Eq. (25) will diminish when the galaxy is appreciably
different from the late type spiral studied in this paper. For example
if the disc deviates from exponential, when there is a substantial
bulge, or when the gas content is much larger. 
It is not expected that medium differences in gas content will play
an important role because as we already saw, the bar develops just
as well with or without gas.
A more general discussion, mainly relating to other stability criteria
will be given in Sect. 9.

As described above, before the bar develops the disc exhibits some
strong spiral features. These features are transient and as a
consequence the disc heats up, possibly aided by additional
scattering on gas accumulations or clouds. When a bar is present
the surrounding disc is heated by emanating arms of variable strength.
This heating leads to an increased disc thickness which is indicated
by the arrows towards larger $z_0$ values in Fig. 11.
As can be seen, when a disc is more unstable the heating and 
thickening is larger, a result which does not come as a surprise.
Noteworthy is the fact that in general when the disc becomes
thicker it remains approximately equally thick at all radii. This 
even holds for the barred region where the thickness is rigorously
calculated as the azimuthally averaged rms z-height and using
$z_0 = 1.10\; z_{\rm rms}$. Only when a bulge develops from gas
transported inwards the thickness remains small or even becomes
smaller in the bulge region.
\subsection{Dead or alive}
The present simulations are performed with a rigid, ``dead'', bulge
and dark halo potential. What might change when instead, the bulge
and halo would be composed of a collection of particles?
To answer this question one has to distinguish between two aspects:
%\smallskip
\begin{flushleft}
\begin{tabular}{lll}
\quad A) & Bar formation & \quad\quad and\\
\quad B) & Bar persistence & \\
\end{tabular}
\end{flushleft}
%\smallskip\noindent
Let's first consider item A. In first instance the development
of the bar depends on the potential of the halo and bulge. 
A more massive halo or bulge increases the potential energy and 
stabilizes against bar formation. For the bulge there is an 
additional effect. When the bulge becomes more massive
an ILR is forced by the higher circular speed near the centre which
frustrates the bar formation (Toomre 1981; Sellwood \& Evans 2001).
These effects are the result of a deeper, azimuthally symmetric,
potential well and consequently changing from a dead to a life
mass distribution will not alter the formation of the bar and the
stability criterion of Eq. (25) remains valid. In second instance,
however, during the process of bar formation there might be
a difference if the halo and/or bulge were alive. A life component
allows angular momentum transfer more easily and resonances between
disc and halo might occur. 
Because the maximum disc case ($\vmd$ = 180 \kms) has a relatively
light halo it is feasible to perform a numerical calculation with
a life dark halo composed of not too massive particles. 
This has actually been done, using 70k halo plus bulge, 70k disc, and
16k gas particles. The result is nearly identical to the case
with a rigid halo and bulge as presented in the bottom row of Fig.~8,
both during and at the end of the simulation after 1.5 Gyrs.
This proves that for at least this situation a dynamical interaction
between an emerging bar with a surrounding life component is not
important.

Bar persistence (item B) is a different story. When an already barred
galaxy is embedded in a life, non rotating isotropic dark halo, the 
bar is slowed down by dynamical friction (Debattista \& Sellwood 1998).
Eventually the bar will then dissolve. This slowing down is severe
when the galaxy is considerably sub maximum and is nearly absent
when close to a maximum disc situation. The initial situation of 
Debattista \& Sellwood is, of course, rather specific and in reality a dark
halo will acquire some rotation during galaxy formation
(Tremaine \& Ostriker 1999). According as a halo rotates faster,
the dynamical friction process is less severe and so in reality
the bar dissolution will be less strong than Debattista \& Sellwood 
claim. Nevertheless, if the simulations were to be performed with a life
halo there would be a dynamical influence on the persistence of the bar. 
Fortunately, as demonstrated above, bars which develope in a more
sub maximum disc situation tend to be smaller. Such bars then reside
in a region where the disc density is still larger than the dark
halo density. Consequently the slowing down mechanism will
only be slightly active if
the halo were life. The mechanism of Debattista \& Sellwood
is important for sub maximum discs with large bars, 
a situation that maybe never occurs in reality. 

A life bulge may, just as a life halo, also influence the persistence
of an already existing bar by dynamical friction. In addition one might
imagine that a small bulge becomes aligned with the bar,
enforcing its presence.
Concerning bulges, it should be kept in mind that the 
investigated galaxy has only a small bulge; the surface density of
the bulge is larger than that of the disc for radii less than
0.9 kpc. The bars which develope are generally much larger than
that extent, and any effect of a life bulge on the bar is therefore
minor. For an earlier type galaxy with larger bulge the
situation might be different.

It would be, nevertheless, interesting to investigate numerically
the effects of a life bulge and halo on the persistence of the bar. 
But that is less simple than it might seem. At first, realistic
and preferably slightly rotating halos and bulges have to be
created. Secondly, halos are massive and large numbers of particles
have to be used. If, instead, one replaces the halo (and bulge)
with less and more massive particles inevitably stochastic heating
of the bar and surrounding disc will occur. The bar then dissolves
by a numerical effect and drawing conclusions is difficult.  

\section{The state of the gas}
\subsection{Rain-clouds}
A good assessment of the state of the gas can be obtained by
plotting the position for every SPH particle in a temperature --
density diagram. Two of such diagnostic plots are shown in Fig. 12
for two exemplary cases: $\vmd$ = 126 \kmss with $z_0$ = 675 pc 
and $\vmd$ = 165 \kmss with $z_0$ = 900 pc.
In all the simulations a typical ``rain-cloud'' structure develops.
Most of the gas resides around 10$^4$~K which is the equilibrium
between FUV heating and the practical upper limit set to the temperature
by the cooling function. In circumstances of high density and/or
low FUV flux gas may cool and starts to rain down from the cloud.
The solid line indicates the Jeans instability threshold. If a 
gas particle falls below this line and remains there for a period
greater than $t_{\rm span}$, half its mass is converted into a stellar
particle and the other half into a hot SN particle. This rain-cloud
structure represents a rather uneasy and dynamical equilibrium state.
If, for example, the FUV flux is removed, nearly all the gas falls
to low temperatures within two time steps. Only at low densities gas
may remain warm by the CR heating. Not shown in the diagnostic plot is
the hot SN gas. This gas has a temperature around 1.25~10$^7$~K
(for $\epssn$ = 25\%) and densities in the range of 0.002 to
0.2 \msol\ ${\rm pc}^{-3}$. If the hot SN phase is switched off the gas
particles generally move back to the 10$^4$~K phase very quickly.
When a bar develops, as for $\vmd$ = 165 \kmss (bottom panel of
Fig. 12) a typical extension on the 10$^4$~K cloud towards high 
density appears. In such cases inwards transport of gas generates
numerous young stars of which the heating outweighs the intrinsic cooling. 

%Fig12
\begin{figure}
\resizebox{\hsize}{!}{\includegraphics{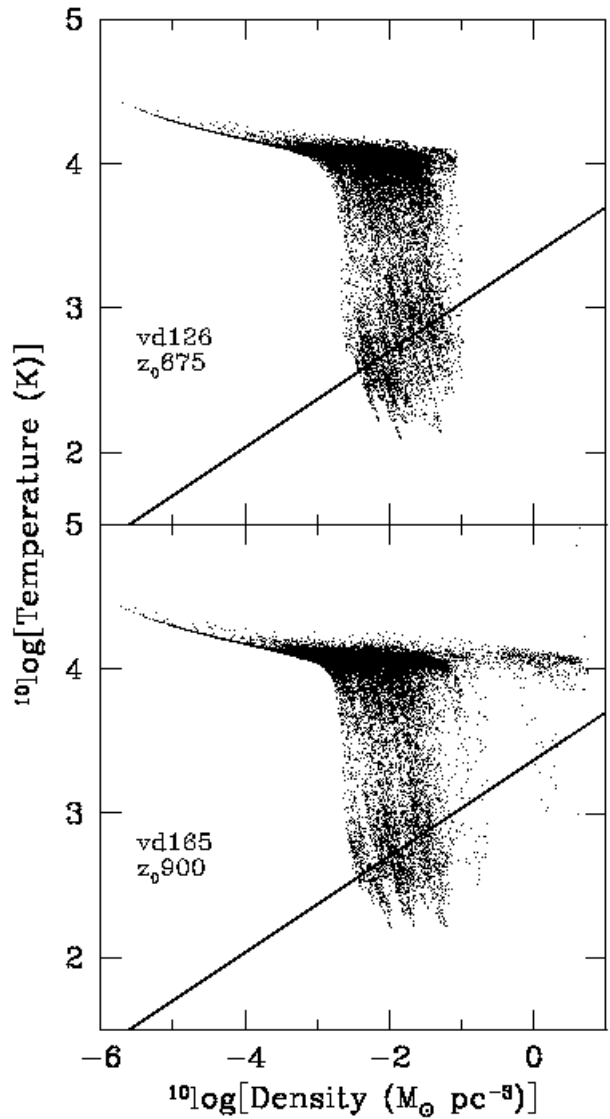}}
\caption[]{
Temperature versus density of the gas for two typical simulations;
$\vmd$ = 126 \kmss with $z_0$ = 675 pc (top panel) and
$\vmd$ = 165 \kmss with $z_0$ = 900 pc (bottom panel). Each dot
represents one gas (SPH) particle. The structure is an equilibrium
between intrinsic cooling and heating by the FUV field. Once a gas
particle falls below the Jeans instability line (full drawn line)
and remains there for a certain time span, it will form a stellar
particle. The extension at 10$^4$~K towards high density in the
bottom panel is typical for gas inside a bar. 
}
\end{figure}

Four temperature ranges have been defined: cold for $T <$ 1000~K,
luke warm for 1000~K $< T < $ 8000~K, warm for 8000~K $< T <$
10$^5$~K, and hot (SN) for $T >$ 10$^5$~K.
For the top panel in Fig. 12 with $\vmd$ = 126 \kmss the total
amount of gas is 18.7~10$^9$ $M_{\sun}$ of which 13.1~10$^9$ $M_{\sun}$ 
resides in the stellar disc region, where $R < $ 25 kpc. Of this amount,
18.5\%, 25.9\%, 54.8\%, and 0.9\% is in the cold, luke warm, warm,
and hot phase respectively. In case of $\vmd$ = 165 \kms, given
in the bottom panel of Fig. 12 the total amount of gas is 17.9~10$^9$
$M_{\sun}$ of which 12.1~10$^9$ $M_{\sun}$ is at $R <$ 25 kpc.
Division of that last amount into the cold to hot phases
gives fractions of 15.8\%, 20.6\%, 62.6\%, and 1.1\%.

There is a considerable amount of luke warm gas; in most cases
even more than the amount of cold gas. Even if one were to increase
the cold fraction by increasing the $f_c$ parameter (Eq.~8) then still
the same amount of luke warm gas remains. This fact is
caused by the abovementioned uneasy and dynamical equilibrium state
of the gas. SPH particles are continuously dropping out of
the rain-cloud by natural cooling and are pushed up there
again by heating. In a manner, this contradicts the proposed notion
of a two phase ISM (Field et al. 1969) or three phase ISM
(McKee \& Ostriker 1977). Although artificially implemented the
hot SN state of the gas can be considered as the hot phase of
this three phase ISM model.

\subsection{The ratio of cold to warm \hi}
The whole process of cloud collapse, division in smaller
fractions, further collapse, star formation, bubble formation, etc, is
extremely complex and not well understood. But anyway, if at a region
gas cools and starts to collapse it takes a certain amount of time for
the first stars to appear. In the present simulations this is implemented
by requiring a certain time span $t_{\rm span}$ (Eq. 8) the gas has
to remain Jeans unstable and thus cold before it may form stars.
The longer $t_{\rm span}$, the more cold gas is present and one
would like to fix this parameter by comparing simulated warm to
cold gas ratios with actually observed ratios in spiral galaxies.
Moreover, the effect of the amount of SN feedback appears to be slightly
related to the amount of cold gas and so related to $t_{\rm span}$.

Unfortunately, observations to determine this ratio are difficult
and the ensuing analysis is ambiguous. In our local Galactic
neighbourhood it seems that cold and warm \hi are evenly divided
(Kulkarni \& Heiles 1987, Payne et al. 1983), though there appears
to be luke warm gas (Kalberla et al. 1985) as well, which
considerably complicates the analysis. For external galaxies the
presence of cold gas has been established (Braun 1997, Young \&
Lo 1996) yet the amount is undetermined. Comparing galaxies,
Dickey \& Brinks (1993) demonstrate that the ratio of cold
to warm gas may differ considerably from system to system yet
its absolute value depends on the unknown temperature (or temperature
range) of the cold gas. Finally I came to conclude that, as for
now, the observations allow a fraction of cold \hi gas anywhere
between 15 and 80\%.
This is all rather disappointing and means that
not much more can be done than quoting gas ratios for the various
simulations.

There are, nevertheless, two interesting facts predicted by the
simulations. At first the presence of a sizable amount of luke
warm gas. Secondly, a large cold \hi gas fraction leads to excessive
clumping and giant SF regions. If the ratio of cold to warm \hi
does appear to be large in an actual galaxy, an additional 
stabilization mechanism is needed for the cold gas. In this
respect magnetic fields might be a good candidate. 

\subsection{\hi, H$_2$, and dust}
An important result of the research by Kennicutt (1989) is that the
Schmidt law holds for the total (H$_2$ $+$ \hi) gas surface density.
This means that the H$_2$ in the inner regions of galaxies actively
participates in generating instabilities and the ensuing star formation.
As for most late type galaxies, also in the inner regions of NGC 628
there is a gradual increase of the H$_2$/\hi ratio (see Fig. 2).
Following the result of Kennicutt, the total gas should and has been
included in the simulations.
But there is a problem. H$_2$ is in a dense region where
according to the Schmidt law there is a lot of SF and consequently
a high FUV field. That field should quickly dissociate the existing H$_2$.
In real galaxies this does not happen and thus another process
is needed to maintain the H$_2$ which is not incorporated
in the simulations.

Above a certain density threshold the atomic gas becomes molecular.
This threshold depends on the pressure, the radiation field, and
metals plus dust. Especially dust is needed to shield the molecules
from photodestruction (Elmegreen 1993). In addition dust allows
H$_2$ to form on grain surfaces (Hollenbach \& Salpeter 1971). Therefore dust
explains, at least for a large part, why H$_2$ can exist in the presence
of the high FUV field. But this also means that at positions with
large H$_2$ column densities there has to be more dust than at other
positions. This is in agreement with the finding of Huizinga \&
van Albada (1992) that absorption in galaxies mainly takes place
in the inner regions.

%Fig13
\begin{figure}
\resizebox{\hsize}{!}{\includegraphics{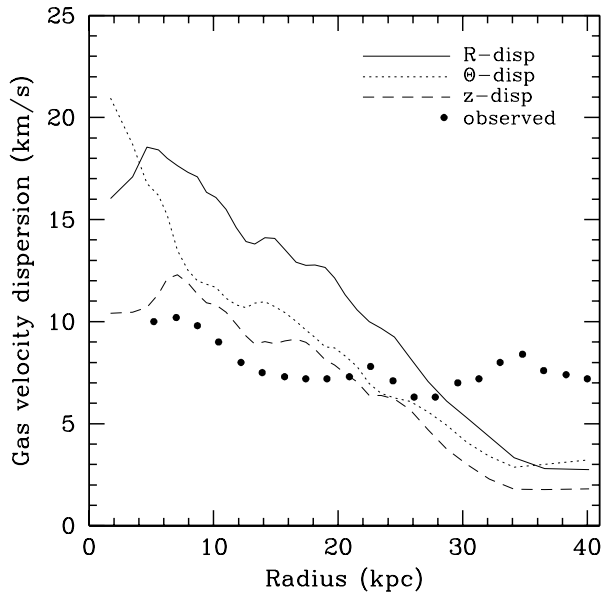}}
\caption[]{
Velocity dispersion of the gas in the three spatial directions
as a function of radius after 1.5 Gyrs of simulation. The dots
give the observed z-velocity dispersion in NGC 628 which is,
inside the stellar disc region ($R < $ 25 kpc), in good agreement
with the simulated values.
}
\end{figure}

Remains the problem that in the simulations the H$_2$ 
is simply replaced with additional
\hi gas. As noted in the beginning of this subsection, that has to
be done to be able to follow the Schmidt law. Yet at the central
regions of a real galaxy heating and cooling might proceed differently
compared to the
\hi mechanism. It will likely proceed at lower temperature regimes
and FUV flux will be attenuated by the extra dust.
Nevertheless one does not expect heating and cooling to be
substantially different because the same metal ions responsible
for the cooling (C$^+$ and O$^{++}$) are present.
Compared to dynamical time scales SF is fast and in a practical
numerical sense always takes place in a few timesteps.
Considering this, and the presence of only
a moderate amount of H$_2$ (8.5\%), it seems reasonable that the
replacement of H$_2$ by \hi in the simulations will not result
in errors on scales larger than the employed spatial and time resolution.

\subsection{The velocity dispersion of the gas}
If there is no SN feedback but only FUV heating the resulting gas
velocity dispersion is in the range of 4 to 7 \kmss in the stellar
disc region. When SNs are included the gas can be stirred up to
higher dispersions as was noted by Gerritsen (1997). For a simulation
with $\vmd$ = 126 \kms, $z_0$ = 675 pc using 40k gas particles
(see next section) the resulting velocity dispersion at
$t =$ 1.5~10$^9$ years is plotted in Fig. 13. By the way, for 16k
gas particles the result is identical. One can see in Fig. 13 that the
velocity dispersion decreases when going to larger radii. This is
caused by less star formation taking place further out in the disc
and consequently there is less SN action, being less able to stir up
the gas. What is interesting is that the radial velocity dispersion
is significantly larger than the dispersion in the other directions,
similar to the stellar velocity dispersions.
The actually observed gas z-velocity
dispersion in NGC 628 (Kamphuis 1993, van der Hulst 1996) is also
given in Fig. 13. There is a good agreement between the
observations and simulations, which suggests that the employed mechanism of
SN feedback quite accurately simulates the processes taking place in
reality.

Actually matters are more complicated than suggested in the
paragraph above. At first because the hydrodynamics is approximated
by a finite resolution numerical scheme. Secondly because the \hi
gas in a real galaxy is not a continuous medium, but a hierarchical
ensemble of clouds. This ensemble of clouds cannot be considered
as a fluid in equilibrium. The clouds and thus the \hi can be heated
by gravitational scattering. In that case a velocity dispersion is
predicted of 5 to 7 \kmss independent of cloud mass
(Jog \& Ostriker 1988, Gammie et al. 1991). Additional heating can,
of course, take place through kinetic energy input by SNs.
It is unclear if, and how accurately the numerical scheme is able to
describe the detailed process of heating for an ensemble of clouds.
An indication that there is indeed such a shortcoming is the
simulated low dispersion values at radii beyond 30 kpc: 2 to 3 \kmss
is simulated while 5 to 7 \kmss is predicted and observed.
Obviously the simulated gas is not discrete enough or too viscous.

%Fig14
\begin{figure*}
\resizebox{\hsize}{!}{\includegraphics{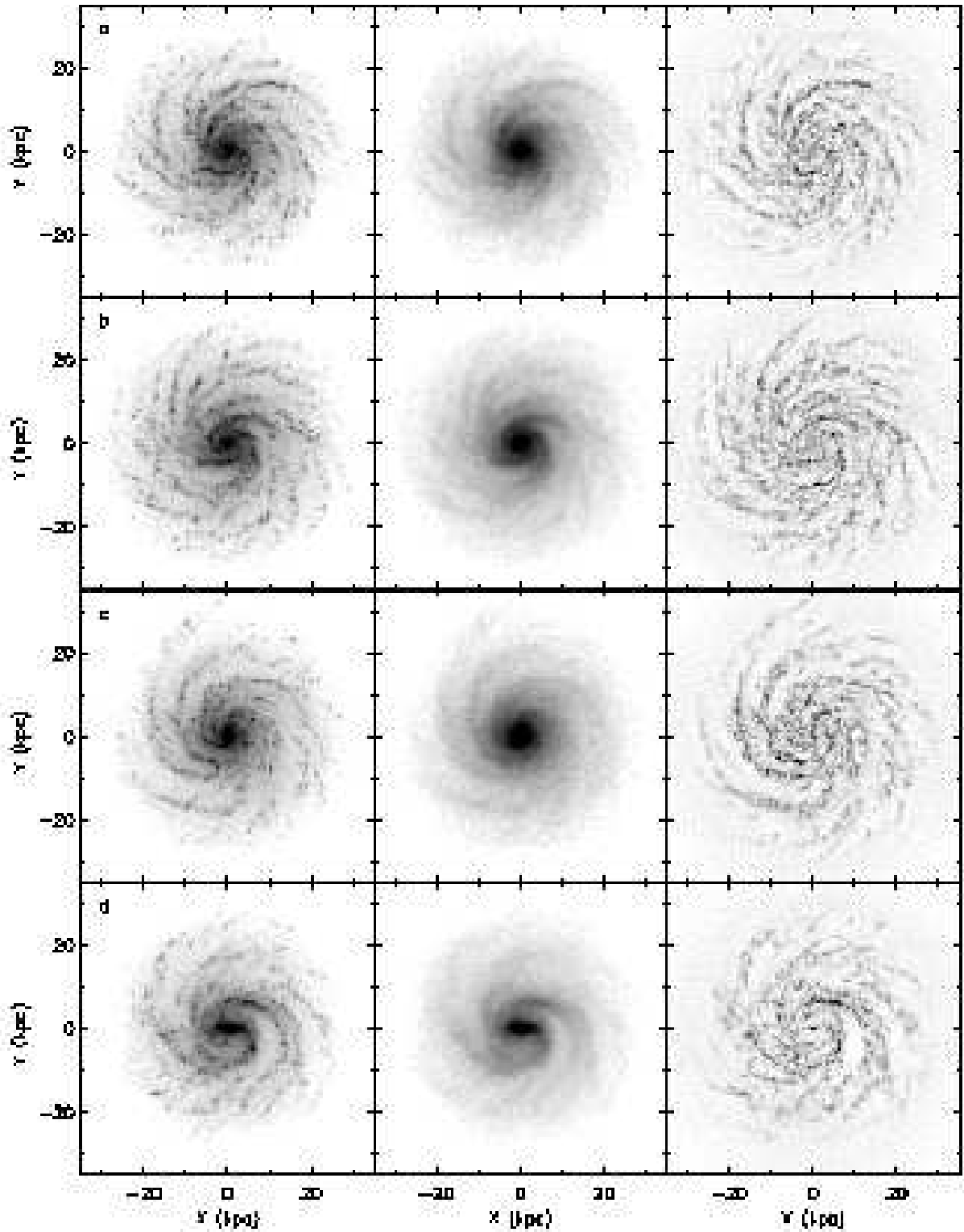}}
\caption[]{
{\bf{a to d}}
The resulting structure of the galaxy using 40k gas and 150k stellar
particles. {\bf a} For $\vmd$ = 126 \kms, $z_0$ = 675 pc, at
$t=$ 1.5 Gyrs. {\bf b} As {\bf a} but now at $t=$ 3 Gyrs.
{\bf c} For $\vmd$ = 150 \kms, $z_0$ = 900 pc at $t=$ 1.5 Gyrs.
{\bf d} As {\bf c} but for $z_0$ = 675 pc.
More gas particles increases the resolution which results in
a more detailed spiral structure specially in the inner regions.
}
\end{figure*}

\section{Spiral structure}
\subsection{40k gas simulations}
In order to make a more detailed
analysis of spiral structure it appeared beneficial to increase
the number of gas particles from 16.000 to 40.000, giving a gas mass
per SPH particle of 4.7~10$^5$ $M_{\sun}$. Simulations with the
increased number of gas particles show lots more details and structure.
To illustrate this, in Fig. 14 the resulting structure is displayed
for four cases which have been investigated for various reasons.
The top two rows are for a galaxy with maximum disc rotation of
126 \kms, $z_0$ = 675 pc, $\epssn$ = 25\% and adding 4 $\msolperyr$
of new gas. Fig. 15a is after a simulation time of 1.5 Gyrs and,
except for the larger particle number, is  equal to the situation
in Fig. 5c. Comparing the two one can notice that especially in
the inner regions more gas particles gives a more pronounced spiral
structure. In the outer regions the spiral arms are more continuous,
less flocculent, and contain fewer star formation spots. Overall,
the general appearance of the mass and gas distribution and of the 
spiral structure has not changed by using more gas particles.
This implies that medium and large scale structures which are generated
do not depend on the employed particle number. A quantitative
analysis of structures can thus be made and the result should
be generally valid.

In Fig. 14b the simulation of Fig. 14a has been continued untill 3 Gyrs
to investigate disc heating and the effect of an increased young
population of stars. The lower two rows of Fig. 14 are for a
galaxy with maximum disc rotation of 150 \kms, $z_0$ = 900 pc
(14c) and $z_0$ = 675 pc (14d) at $t =$ 1.5 Gyrs. These two simulations
are on opposite sides of the bar instability threshold for that
maximum disc rotation. Fig. 14c is, except for the larger particle
number, equal to Fig. 8b and the same conclusions apply as for the
comparison of the $\vmd$ = 126 \kmss discs, given above.

\subsection{Stellar heating}
A long simulation of a stable disc provided the possibility to
investigate the evolution of the stellar velocity dispersion.
New stars are formed at a rate of approximately 4 $\msolperyr$
with an original velocity dispersion equal to that of the gas.
Over a time span of a few Gyrs these young stars will acquire
an ever larger velocity dispersion.

The proven stable disc with $\vmd$ = 126 \kmss and $z_0$ = 675 pc was
simulated for 3 Gyrs using 150k stellar, and 40k gas particles.
For the whole period gas was
added to the system at a rate of 4 $\msolperyr$ to compensate
for the gas consumption by star formation. At $t =$ 3 Gyrs, the
situation displayed in Fig. 14b, the then available stars were
divided in five age groups. The youngest stars have an age $<$
0.75 Gyrs, there are three groups with ages between 0.75 to 1.5 Gyrs,
1.5 to 2.25 Gyrs, and 2.25 to 3 Gyrs, and there is a group of old
stars present from the onset of the simulation, thus having
an age $>$ 3 Gyrs. At radii of 5, 10, 15, and 20 kpc the velocity
dispersion in both the radial and vertical direction was determined
of these age groups. The result is given as a function of the
square root of the age of each group in Fig. 15. In this way the evolution
of the dispersion can be monitored at each radial position.
As can been seen, the increase in dispersion is nicely consistent
with a square root of age functionality. After 3 Gyrs the dispersion
of the old population has not yet been reached. If one extrapolates
the data points to larger ages, then on average, after $\sim$
6.5 Gyrs the dispersion of the old, equilibrium population
of stars will be reached.

An overview of the rates and the mechanisms of disc heating is given
by Binney \& Tremaine (1987, p. 470 - 484) and by Lacey (1991).
In short, the collisionless stellar ensemble of a stellar disc can
be heated by three principal mechanisms. Firstly, heating by
molecular clouds (Spitzer \& Schwarzschild 1953, Villumsen 1985),
which generates heating $\propto t^{0.25}$. Secondly, heating
by transient spiral features (Carlberg \& Sellwood 1985) which is more
efficient and heats $\propto t^{0.5}$ but molecular clouds are
needed in addition to generate vertical heating. A third mechanism
is heating by the infall of small satellite galaxies. Stellar heating
as observed in the solar neighbourhood indicates that the heating,
or increase in velocity dispersion, proceeds proportional
to $t^{0.5}$ (Wielen 1977) so that the more effective spiral arm
heating is needed to explain this observation.

The simulation of Fig. 14a and b actually includes, both heating by
simulated molecular clouds (conglomerates of gas particles), and
heating by transient spiral arms. It should therefore give a reasonable
description of heating as occurring in real galactic discs. As
witnessed in Fig. 15 the simulated heating is indeed consistent with
a proportionality $\propto t^{0.5}$, as observed in our Galaxy and
predicted by spiral arm heating. However, as mentioned in
Sect. 4.4 the simulation is only collisionless for $R \la 2h$.
Therefore, for radii larger than $\sim$ 9 kpc the simulations have
additional heating by particle scattering. This will generate
heating time scales being shorter than for real galaxies, certainly
at the radial positions of 15 and 20 kpc.

%Fig15
\begin{figure}
\resizebox{\hsize}{!}{\includegraphics{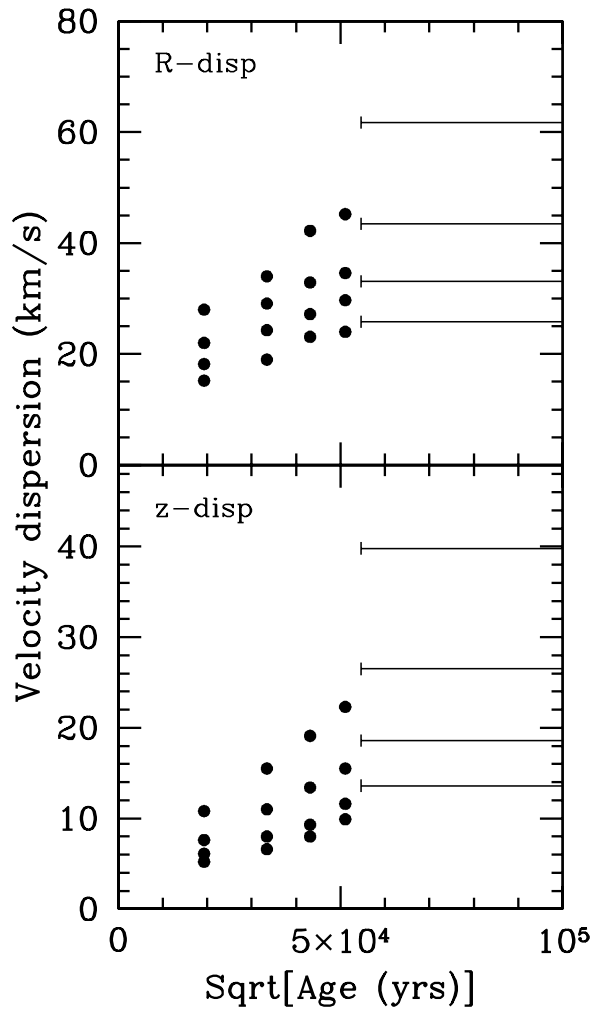}}
\caption[]{
The time evolution of the stellar velocity dispersion in the
radial direction (top) and vertical direction (bottom panel) derived
from the simulation presented in Fig. 14b. Stellar particles are
divided up into five age bins at four radial positions of
5, 10, 15, and 20 kpc. For each bin at each radius the dispersion
is calculated. The old stars, present from the onset of the
simulation (age $>$ 3 Gyrs) are indicated by the bars. At each
age bin a larger dispersion is for a smaller radius. It appears
that the stellar heating is consistent with being proportional
to $t^{0.5}$.
}
\end{figure}

\subsection{On what does the amount of spiral structure depend?}
My original perception concerning the creation of spiral structure
was that its origin lies in an underlying young and cold stellar fraction.
To investigate this assumption, for two stable discs ($\vmd$ = 126 \kms,
$z_0$ = 675 pc and $\vmd$ = 150 \kmss with $z_0$ = 900 pc) long
simulations using 40k gas particles have been performed until
3~10$^9$ years. During this time young stars have been created
at a rate of $\sim$ 4 $\msolperyr$ at the position and with the
kinematics of the gas layer. The result was disappointing. As can be
seen in Fig. 14a and b, the spiral structure might be marginally
stronger at 3 Gyrs compared to that after 1.5 Gyrs. However, if the
whole time sequence for this $\vmd$ = 126 \kmss run is inspected then
there are no gradual changes any more after $\sim$ 1 Gyrs. There are
fluctuations in the appearance and strength of the spiral arms over
periods of $\sim$ 0.2 Gyrs which is illustrated as the difference
between Fig. 14a and Fig. 14b. The same conclusion applies for the
$\vmd$ = 150 \kms, $z_0$ = 900 pc case (Fig. 14c). If the addition
of gas is stopped after 1.5 Gyrs, the SFR decreases to $\sim$
2 $\msolperyr$ at $t =$ 3 Gyrs and the spiral structure slowly
fades away.

In retrospect this failure of a young population to generate a more
pronounced spiral structure can be understood. For example, the
$\vmd$ = 126 \kmss disc has a total stellar mass of 45.5~10$^9$
$M_{\sun}$. During the simulations cold stars are added at a rate
of 4 $\msolperyr$. After 1 Gyrs the cold stellar fraction then
amounts to 4~10$^9$ $M_{\sun}$, and after 3 Gyrs to 12~10$^9$ $M_{\sun}$
which is approximately 20\% of the then available total stellar mass.
However, the stars which were formed first are then not young
and cold any more because of stellar heating. As can be seen in
Fig. 15, after 3 Gyrs, the stars with that age have increased their
velocity dispersion to more than half of that of the old population.
Considering the numbers and heating rates, the mass fraction of a
reasonably cold stellar population will not surpass the amount
of 20\%. Such a fraction might locally destabilize the disc to some
extent but will never become a dominant mechanism.

The total mass of a disc with $\vmd$ = 150 \kmss is 64.5~10$^9$
$M_{\sun}$. In that case for a SFR of 4 $\msolperyr$ the mass fraction
of cold stars is approximately 10\%, being a factor two smaller
than for the $\vmd$ = 126 \kmss disc. For a more massive disc, with
the same gas content it would thus be even more difficult to create
instabilities by the cold stellar population process, and vice versa,
of course. Some additional experiments have been carried out by adding
large quantities of gas during the beginning of the simulations. As
a result the SFR goes up and a larger cold stellar fraction is generated.
The spiral structure then becomes more pronounced, but subsides to its
usual state once the extra gas flow is stopped. In these cases
it is not clear whether the more pronounced spiral structure is
generated by the larger cold stellar fraction or by the larger
amount of gas which is present.

Since an increased cold stellar fraction barely enhances the spiral
structure, the influence of other parameters has been investigated.
It was apparent from the investigation of Sect. 5 that spiral
structure is mainly caused by shearing gas structures and an increased
fraction of cold gas might enhance such a process. For the situation
of Fig. 14c, $\vmd$ = 150 \kmss and $z_0$ = 900 pc, the collapse
factor $f_c$ (Eq. 8) has been increased by a factor four in
combination with or without a change of the supernova efficiency.
A larger collapse factor produces more cold gas, which is colder
and has densities up to a factor of 10 larger. The gas arms are thinner
and more threadlike and tend to form gas clumps which cause strong
star formation regions. For a combination of $t_{\rm span} =
2 \times t_{\rm ff}$ and a low SN efficiency of $\epssn$ = 10\% the
image is like the top rows of Fig. 5 and not like any actual
galaxy. Consequently, the attempt to make more massive gas concentrations
by increasing the cold gas fraction and lowering the SN dispersal
mechanism is successful, but produces the familiar clumps and does
not enhance a global spiral pattern. Letting $\epssn$ remain at 25\%
or making it larger in combination with a larger cold gas fraction
does generate the more threadlike gas structure, but does not
create larger or significantly smaller spiral disturbances in the
underlying stellar density structure. What this investigation makes clear
is that the generated spiral structure as it is in Fig. 14c is only
weakly dependent on the exact value of the gas parameters.
Putting it in other words, a significant change of cold gas
fraction or supernova efficiency does not lead to large changes
in the spiral structure morphology.

The standard cooling function applied to the situation of NGC 628
leads to a slightly large value of the observable SFR, when compared
with what is actually observed. Such a large SFR had the advantage
of quickly generating an appreciable amount of young and cold stars
of which the effect on the emerging spiral structure could be investigated.
Since such effects are demonstrated to be minor, the cooling function
can be changed so as to generate a lower SFR. This has been investigated for
the present case ($\vmd$ = 150 \kms, $z_0$ = 900 pc) by lowering
the SFR to $\sim$ 2$\frac{1}{2}$ $\msolperyr$ and adding only
2 $\msolperyr$ of gas constantly. It appears that there are no
noticeable differences in the generated spiral structure compared to
a situation with a larger SFR.
From this follows that spiral structure at low gas suppletion rates
can be sustained longer if the cooling of the gas is less efficient.

One can conclude that small and medium scale spiral structure as
displayed in Fig. 14a to c can be generated in an isolated spiral
disc, for a large range of gas parameters. The principle factor
determining the strength of the spiral structure is the surface
density of the gaslayer. However, it is not possible to create
autonomously a large scale spiral density wave. If the disc is thinner
and colder and crosses the threshold of bar instability the bar
can then drive an $m = 2$ spiral structure as demonstrated in Fig. 14d.

\subsection{A quantitative analysis of spiral structure}

%Fig16
\begin{figure}
\resizebox{\hsize}{!}{\includegraphics{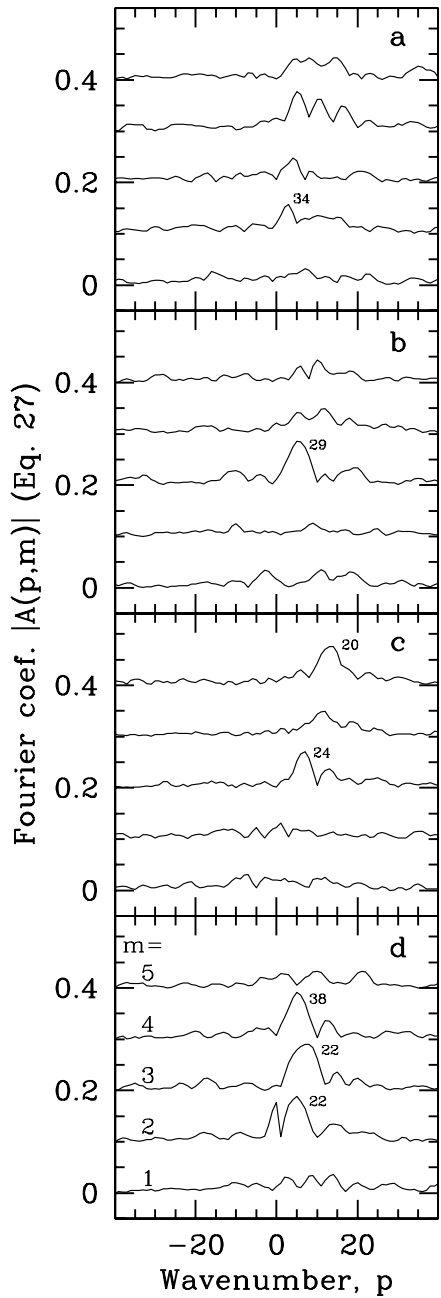}}
\caption[]{
{\bf{a to d}}
Amplitude of the Fourier coefficients $A(p,m)$ of the V-band
images in Fig. 14a to d using global logarithmic spirals as base
function (Eq. 27). For each panel the lines from bottom to top
are for $m$ = 1 to 5 and are offset by 0.1 with respect to each
other. The belonging pitch angle is indicated by the numbers for
a few features. At panel {\bf d} the signal at $p=0, m=2$
originates from the bar of Fig. 14d.
}
\end{figure}

One would like to compare the spiral structure produced by the
simulations with the actual spiral structure in real galaxies in
some quantitative way. For example, the number of spiral arms, their
shape, and strength. Astonishingly only a small amount of research
has been carried out in this respect and a comparison of simulations
with observations is only marginally possible.

The shapes of spiral arms are generally well represented by a
logarithmic functionality (Danver 1942, Kennicutt 1981) as
\begin{equation}
r = r_0 {\rm e}^{{\rm tan}(i_p)\; \Theta},
\end{equation}
where $r$ is the position on the arms with belonging angle $\Theta$ and
$i_p$ is the pitch angle. Rewriting Eq. (26) gives
${\rm ln}(r) \propto {\rm tan}(i_p)\; \Theta$. When the density
structure of a (face-on) galaxy is rebinned to a $({\rm ln}(r), \Theta)$
plane, logarithmic spirals show up as straight lines. Derivation of the
pitch angles for each arm is straightforward and the number of
crossings of one feature over the $2 \pi$ long $\Theta$ axis gives the
$m$-mode of the feature. On various occasions simulations have been
displayed in such a plane and the simulated arms are always well
represented by a logarithmic shape. However, $({\rm ln}(r), \Theta)$
planes are not easily comparable because there is in fact too much
and too dispersed information. One would like to reduce the information
into less and more informative numbers.

One method to do this has been explored by Consid\`ere \& Athanassoula
(1982, 1988). It involves a Fourier transform of the image of the
galaxy into a superposition of global $m$-armed logarithmic spirals
following Kalnajs (1975). A two dimensional grid $(i,j)$ is transformed
into Fourier coefficients $A(p,m)$ by
\begin{equation}
A(p,m) = \frac{
 {\sum}_i {\sum}_j w_{ij} {\rm exp}
[ -i( p\; {\rm ln}(r_{ij}) + m {\Theta}_{ij})]
}{
{\sum}_i {\sum}_j w_{ij}
}
,
\end{equation}
where $p$ is the wavenumber, $m$ the mode and $w_{ij}$ the value at
gridpoint $i,j$. The pitch angle $i_p$ is given by
${\rm tan}(i_p) = -m/p$ at the wavenumber where $|A(p,m)|$ has its
largest amplitude.
Consid\`ere \& Athanassoula (1988) present the calculated Fourier
coefficients in their Fig. 5 for 16 spiral galaxies, among which
are a number known to have a well developed grand design
structure. For most of the galaxies the coefficients
$|A(p,m)|$ rarely surpass the level of 0.1. Notable exceptions
are NGC 5247 $(|A(p,m=2)| \sim 0.45)$ and NGC 6946 $(|A(p,m=2)| \sim 0.21)$,
but these two galaxies both have exceptionally large pitch angles
of 26 and 32\degr\ respectively. Consid\`ere \& Athanassoula
claim that the two-armed component is everywhere dominant, but a close
inspection of their data shows that only for 8 out of 16 galaxies
$m=2$ is dominant, while for 6 out of 16 the $m=1$ component
dominates! Surely this indicates that a number of systems have
undergone some kind of interaction. NGC 628 is in the sample
and shows a maximum $|A(p,m=2)|$ of 0.16 with a pitch angle of 16\degr.
In addition there is a peak for $m=3$ at a level of 0.09 with
pitch angle of 28\degr.

For the V-band images of the simulated galaxies in Fig. 14 the Fourier
coefficients for modes one to five have been calculated following
Eq. (27) and are displayed in Fig. 16. In general the peak values do
not exceed the level of 0.1, as for most of the real galaxies, in fact.
Fig. 16 is only one snapshot; if the Fourier coefficients are
monitored over a time span, peaks come and go for modes 2 to 5.
Time scales are typically 0.2 Gyrs and levels are comparable to those
displayed in Fig. 16a to c. This behaviour reflects the episodical
nature of the small and medium scale spiral structure during the simulations.
For Fig. 16d, the barred galaxy, peaks are more constantly present.
The simulations are supposed to represent NGC 628. As can be seen in
Fig. 16, the values of the Fourier coefficients do not reach the
observed value. In addition, simulated pitch angles are somewhat
too large. Although the simulated V-band images reasonably resemble
images of actual galaxies, the observed image of NGC 628 (for example
on page 99  of the Shapley-Ames catalog, Sandage \& Tammann 1981)
shows arms which are longer and less open. This means that the
real galaxy NGC 628 exhibits more a grand design than its simulated
counterpart. A further discussion of this matter
will follow in Sect. 10.

As determined by Kennicutt (1981), pitch angles of late type galaxies
lie in the range between 12 and 30\degr\ and have a median value of 18\degr.
As such, the pitch angles of the simulated galaxies are slightly large
but certainly not unusual. Note that for the galaxies in common to the
samples of Consid\`ere \& Athanassoula and Kennicutt the derived
pitch angles are in good agreement. But this brings me to the question:
how representative are both these samples for an average galaxy?
Probably not representative at all. For both samples, galaxies are
selected at least partly, for having a grand design or otherwise
well delineated spiral structure. Inevitably this produces a severe
bias towards high amplitude spiral signal and probably towards smaller
pitch angles.

%Fig17
\begin{figure}
\resizebox{\hsize}{!}{\includegraphics{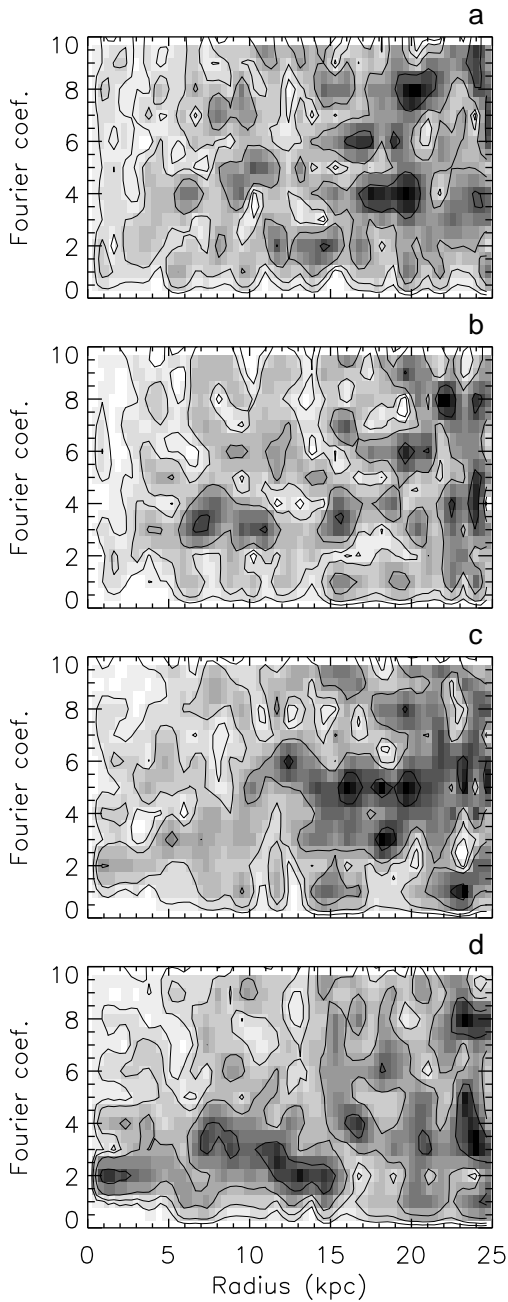}}
\caption[]{
Amplitude of the coefficients of the Fourier sum along
circular annuli as a function of radius for the four V-band
images of Fig. 14a to d. Contour levels are at 5, 10, 20, 40,
and 80\% of the underlying smooth galactic brightness level.
Specific features can easily be related to features in
Fig. 14 and Fig. 16.
}
\end{figure}

An alternative way to analyse spiral structure has been employed by
a.o. Elmegreen et al. (1989, 1992), Rix \& Zaritsky (1995), and Fuchs
\& M\"ollenhoff (1999). According to this method, first the
underlying smooth galactic  distribution is removed. Then the
residual is divided up in a number of circular annuli of which
the brightness $B$ is Fourier analysed with respect to the azimuthal
angle $\phi$:
\begin{equation}
B = {\sum}^{10}_1 c_m {\rm cos}(m({\phi}-{\phi}_m)).
\end{equation}
Following Fuchs \& M\"ollenhoff the coefficients $c_m$ will be displayed
as a function of radius to obtain a direct and quantitative insight
into the development of the multiplicity of the structure. A
determination of the pitch angle of mode $m$ might be done by
plotting the phase ${\phi}_m$ versus ln(radius), but this has
not been investigated presently.

The V-band images of Fig. 14 have been divided up into 35
equally wide annuli ranging from a radius of 0.5 to 25 kpc. After
division by the continuum, the Fourier coefficients of Eq. (28)
were calculated and are displayed in Fig. 17. One may notice
that a good impression can be obtained of the radial functionality
of the multiplicity of the spiral structure. Comparing Fig. 16
and Fig. 17, certain features can be related and put at its proper
radius. For example, in panel {\bf a} the small $m=2$ signal originates
from radii between 11 and 15 kpc, while the $m=4$ signal comes
mainly from positions in the outer regions. In panel {\bf b} the $m=3$
spiral with pitch angle of 29\degr\ can obviously be
recognized in Fig. 17. On the other hand, the small $m=2$ signal
in panel {\bf c} of Fig. 17 at radii $<$ 3 kpc cannot readily be
seen in Fig. 16c. Such a feature appears small in relative units,
but if this disturbance of $\sim$ 15\% is expressed in absolute
values for NGC 628 it amounts to $\sim$ 25 $L^V_{\sun} {\rm pc}^{-2}$,
which is certainly not negligible. Expressed in relative units,
Fig. 17 shows that the brightness fluctuations become larger at
larger radii. In absolute units, however, the signal dominates in the
inner regions. For the barred galaxy in panel {\bf d} it can be noticed that
the $m=2$ inner bar drives the $m=2$ spiral arms further out,
except for the small excursion to $m=3 - 4$ around a radius of 8 kpc.
Of course, during the simulation the spiral structure
changes and consequently also its representation as given in Fig. 17.

Though this is a nice way to display the spiral structure of a galaxy,
there is a major problem. There are hardly any observations which
have been interpreted in this manner and hence a comparison between
simulations and real data is not really possible. Rix \& Zaritsky (1995)
give $c_1$ to $c_6$ values for $K^{\prime}$-band observations of
18 face-on spirals. Their main intent was to investigate the lopsidedness
of galaxies and thus they focussed on the $c_1$ coefficient. The
near infrared data do not extend beyond three scalelengths and the
values of the $c_3$ to $c_6$ coefficients over that radial extent
are relatively small, yet consistent with the values in Fig. 17.
Only Fuchs \& M\"ollenhoff (1999) give a detailed Fourier coefficient
analysis for one galaxy: NGC 1288. Both, the values and radial
functionality of the coefficients show a striking similarity with the
simulations of non-barred galaxies. NGC 1288 is a regular, massive
galaxy with clear spiral structure but has no grand design density
wave permeating its disc. As such it is comparable with the simulations
and the spiral structure shows a comparable morphology.
\section{Swing amplification}
The spiral structure which develops in the present numerical
simulations of an isolated and bar stable galaxy appears to be
generated by swing amplification. This phenomenon where small distortions
can be severely amplified in a differentially rotating (shearing) disc
was discoved by Goldreich \& Lynden-Bell (1965) and the theory has
been elaborated by o.a. Julian \& Toomre (1966), Toomre (1981),
Athanassoula (1984), and Athanassoula et al. (1987). As noted by
Toomre (1981) swing amplification is governed by a conspiracy between
three processes: shear, shaking and self-gravity. Each of these
processes can be specified by a dimensionless parameter: 
$\Gamma$, $Q$, and $X$ which are the shear rate, the familiar
Toomre's $Q$ parameter and as it is sometimes called Toomre's 
$X$ parameter. In principle, swing amplification has been derived and 
quantified for a one component, infinitely thin disc using the 
epicyclic approximation. That is a simplification of reality
which should be kept in mind when comparing results of the theory
with findings for a real or numerical galaxy. 

To investigate whether swing amplification is indeed the cause
of the numerical spiral structure, the theory has been applied
to the galaxies displayed in Fig. 14. The three abovementioned 
parameters have to be calculated for the galactic constitution which
leads again to the problem encountered in Sect. 6.2 of what to
substitute for the rotation and surface density.
As in Sect. 6.2 where two kinds of $Q$ are defined
also here two kinds of $\Gamma$ and $X$ will be defined; pure
stellar (${\Gamma}_{\ast}$ and $X_{\ast}$) given by
\begin{equation}
{\Gamma}_{\ast} = - \frac{R}{{\Omega}_{\ast}}
\frac{d {\Omega}_{\ast}}{ d R},
\end{equation}
\begin{equation}
X_{\ast} = \frac{ {\kappa}_{\ast}^2 R }{ 2 \pi G m {\sigma}_{\ast} },
\end{equation}
and total or testparticle parameters (${\Gamma}_t$ and $X_t$)
given by
\begin{equation}
{\Gamma}_t = - \frac{R}{ {\Omega}_{\rm gas} }
\frac{ d {\Omega}_{\rm gas} }{ d R },
\end{equation}
\begin{equation}
X_t = \frac{ {\kappa}_{\rm gas}^2 R }
{2 \pi G m ({\sigma}_{\ast} + {\sigma}_{\rm gas}) }.
\end{equation}
Here $\Omega$ and $\kappa$ are the orbital and epicyclic frequencies,
$\sigma$ is the surface density and $m$ is the azimuthal multiplicity
of the disturbance which is equal to the number of spiral arms.
For the galaxies of Fig. 14a ($\vmd$ = 126 \kms, $z_0$ = 675 pc, 
$t$ = 1.5 Gyrs) and Fig. 14c ($\vmd$ = 150 \kms, $z_0$ = 900 pc,
$t$ = 1.5 Gyrs) the swing amplification parameters are given as a
function of radius in Fig. 18. As can be seen, there is only a marginal
difference between ${\Gamma}_{\ast}$ and ${\Gamma}_t$. On the 
other hand, the $Q$ and $X$ parameter may differ substantially mainly 
depending on whether the gas surface density is, or is not included.

%Fig18
\begin{figure}
\resizebox{\hsize}{!}{\includegraphics{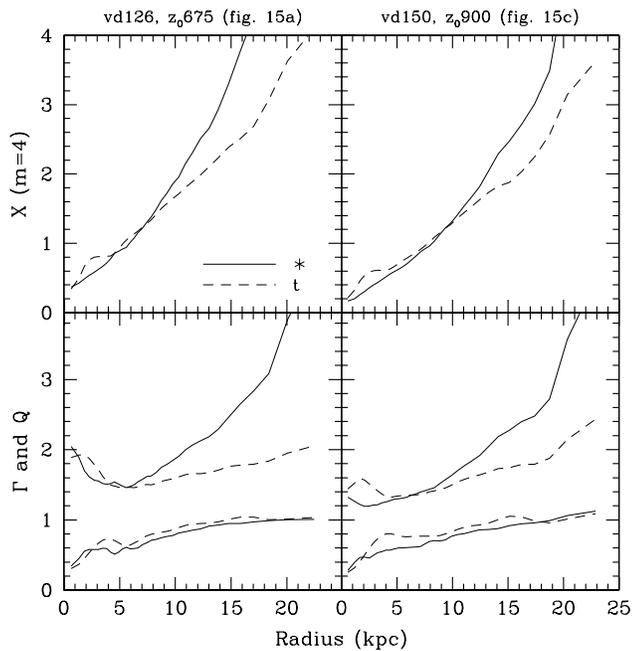}}
\caption[]{
Swing amplification parameters $X(m=4), {\Gamma},$ and $Q$  as a 
function of radius for the galaxies of Fig. 14a and Fig. 14c. These
parameters are defined in two ways ($\ast$: Eq. 22+29+31,
$t$: Eq. 23+30+32) depending on the way the influence of the gas
is included.
}
\end{figure}

Athanassoula (1984) has
calculated the maximum amplifications for a number of different values
of the three dimensionless parameters. By interpolating her Fig. 26,
the amplification can thus be found. But this is a maximum amplification
provided the arrival phase of the distortion is optimum. This means
that for an actual case the amplification may be less, or zero, or the
amplification of a distortion which emerges at a later time with
more optimum arrival phase may approach this maximum limit. When
viewed over a longer time span the spiral pattern will thus be rather
variable just as witnessed in the simulations. Inspection of 
Athanassoula's Fig. 26 establishes some generalities, though.
For a $\Gamma$ value between 0.5 and 1.0 swing amplification
is only effective for $X$ between typically 0.5 and 2.5 and has
the largest amplifications when $X$ lies between 1.0 and 1.5. A large
value of $Q$ dampens the amplification; when $Q \ga 2.0$ the
amplification is $\la$ 3. When $Q$ is low, say $Q \la 1.2$ then
the amplifications may exceed the value of 30.

For the present two galaxies of which the parameters are displayed
in Fig. 18 the maximum amplifications have been derived as a function
of radius. To that aim the pure stellar $({\ast})$ and total ($t$)
curves have simply been averaged using the assumption that the gas
layer and its gravitation will influence the swing amplification process
to some extent. If the amplifications would have been derived for
both curves separately the emerging picture would not be essentially
different. This is because an increasing $Q$ value results in a
decreasing amplification with a factor barely dependent on $\Gamma$
and $X$. A difference between the $X$ curves
mainly appears at large values of $X$ where the amplification
is small anyway. The maximum amplifications for modes 2, 3, 4, 6, and
12 are displayed as a function of radius in Fig. 19.

%Fig19
\begin{figure}
\resizebox{\hsize}{!}{\includegraphics{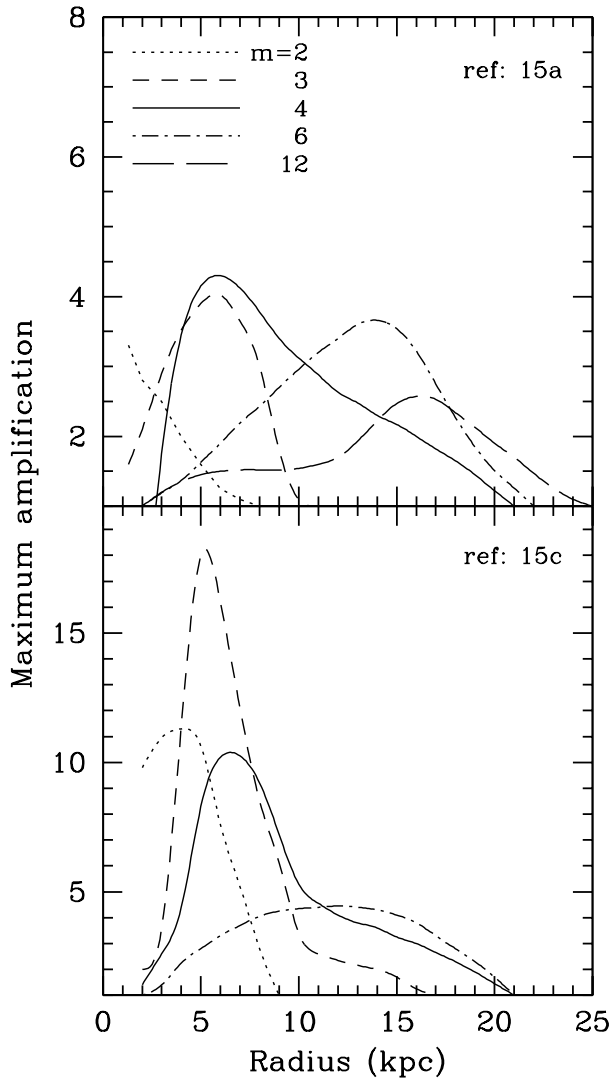}}
\caption[]{
The maximum amplification factors given by swing amplification
theory for the galaxies of Fig. 14a (top) and Fig. 14c (bottom).
Note that at certain radii the amplifier may be active
and equally strong for different arm multiplicities. 
}
\end{figure}
 
One may notice that the smaller $m$-modes ($m$ = 2, 3 ) or the two and
three armed spiral structures allow the largest amplification at
the smaller radii. Spiral structures with more arms will preferentially
develop further out. At certain radii the amplifications for a number
of different modes can approximately be equally large. In practice
that means that the number of arms can, and likely will, change from time 
to time. For the less massive disc ($\vmd$ = 126 \kms, top panel
of Fig. 19) the amplifications reach at most a factor of four
which is mainly caused by the relatively large value of $Q$.
When that simulation is continued to 3 Gyrs as displayed in Fig. 14b
the swing amplification parameters barely change. Consequently also
the amplifications appropriate for that later time are equal to those
already given for $t$ = 1.5 Gyrs. If the disc is more massive
(bottom panel of Fig. 19), $Q$ is smaller and the allowed amplifications
are larger certainly for the smaller $m$-modes. For the same disc
but with a smaller thickness ($z_0$ = 675 pc instead of 900 pc) as in
Fig. 14d the resulting $Q$ parameter (at $t$ = 0) is smaller while
the $X$ and $\Gamma$ parameters remain nearly equal. The emerging
amplification scheme for that thinner disc is then equal to that
in the bottom panel of Fig. 19 except that all the amplifications
have to be multiplied with a factor 2$\frac{1}{2}$ to 3. That results
in values between 30 and 40 for $m=$ 2, 3, and 4 at radii $<$ 
10 kpc. But, as we know a bar develops in that case which might be
related to these allowed large amplifications.
When one compares the amplifications given in Fig. 19 with the
actually emerging modes in Fig. 17 one may note the exceptional 
correspondence. For example, the $m=3$ structure in Fig. 17b between
6 and 11 kpc and the same allowed mode by swing amplification at
exactly the same radii. Even nicer is the correspondence between
Fig. 17c and Fig. 19 bottom panel. In that case the radial functionality
of the appearing structure and the predicted amplifications are in
striking agreement. 

To investigate if the magnitude of the disturbances is consistent
with swing amplification theory, for the case of Fig. 14c the 
initial gas distortions have been compared with the eventual strength
of the stellar mass arms. For example at $R$ = 3 kpc gas mass fluctuations
are $\sim$ 3.5 \smppc2\ and the $m=2$ stellar mass arms have a 
strength of $\sim$ 23.6 \smppc2\ implying an amplification of a 
factor $\sim$ 6.7. The allowed maximum amplification of $m=2$
at $R$ = 3 kpc is $\sim$ 11, in perfect agreement with the simulations.
Likewise, at $R$ = 5 kpc for $m=3$ one has an amplification of $\sim$
4.7 while a maximum of $\sim$ 17 is allowed, and at $R$ = 13 kpc
amplifications are $\sim$ 2.2 while $\sim$ 4.5 is allowed.

One may conclude that the predictions of swing amplification theory and
the results of the numerical simulations are completely consistent.
Not only for the radial positions where the modes preferentially occur,
but also for the actual strength of the spiral arms.

\section{Discussion and conclusions}
\subsection{On grand design}
For a late type bar stable galaxy in isolation it does not
seem possible to have a  grand design spiral density wave pervading
its disc. The spiral structure which does develop is on a medium
and small scale with a multiplicity increasing as a function
of radius. That finding can be explained qualitatively and 
quantitatively by swing amplification; a process which occurs
naturally in a differentially rotating disc. If swing amplification can
explain the structure for a late type galaxy so well, it is likely
that it also applies for other, more earlier types. The question
which then emerges is whether for some galaxy swing amplification
allows a spiral structure with a constant multiplicity over a sizable
range of radii. Such a situation would, in principle, be equal
to a grand design.

To answer the question one could calculate the amplification
factors for a number of observed galaxies and investigate if a
certain mode is dominant over a large radial extent. Such a calculation
has been done by Athanassoula et al. (1987) but unfortunately
they do not present the radial modal functionality. A global
insight can be obtained by considering equations 29 and 30 and Fig. 18.
To make a constant radial multiplicity, the $\Gamma$ parameter
can only be slightly variable and the $X$ parameter needs to be 
nearly constant. For NGC 628 (Fig. 18) that is certainly not the case. 
Making a constant $X$ requires that ${\kappa}^2 R/{\sigma}$ is constant,
while in the mean time the rotation curve should be rather flat. 
That results in the condition that ${\sigma}R \sim$ constant which
is impossible to achieve for an exponential disc $(\sigma \propto
{\rm e}^{-R/h})$. It can, however, not be excluded that for some
rotation curve and photometry a constant $X$ over an extent of,
say, two scalelengths can be achieved.

For a number of well known grand design galaxies the $m=2$ mode is
the dominant one. How can such a situation arise when swing 
amplification is the driving mechanism? For a late type galaxy
(like NGC 628) figures 18 and 19 can again give some insight.
In case of a maximum disc rotation of 126 \kmss (and $z_0$ 
= 675 pc) the amplifications are given in the top panel of Fig. 19.
For radii between 3 to 10 kpc the $m=4$ mode is the largest.
Inspection of Eq. (30) shows that exactly the same $X$ values
are achieved when half the multiplicity $(m)$ is used and a 
double value of the surface density. Then Fig. 19 (top) has also 
the same appearance but the $m=4$ is replaced with $m=2$, while
the amplifications are much larger because $Q$ is smaller.
As a consequence $m=2$ is dominant between 3 and 10 kpc and might
drive a grand design $m=2$ through the rest of the galaxy.
The problem is, however, that the surface density has to be doubled,
which makes the disc massive, with a maximum rotation of 
$\sim$ 180 \kmss and then a bar emerges. So, as soon as a late 
type galaxy reaches the state that $m=2$ is active over a considerable
radial extent and with enough amplitude, its disc is so massive
that instead a bar develops. This explains the results of the 
numerical calculations in this paper, namely that only multiple,
non grand design spirals appear, or a bar when the disc is more massive.
Unless, of course, one explicitly prevents the
development of a bar (Elmegreen \& Thomasson 1993). Then a grand
design $m=2$ may exist in a hypothetical galaxy. 

For late type galaxies it seems impossible to create a grand
design and for other disc like galaxies it seems to be very
difficult. How does nature then make the well known beautiful
spiral galaxies which adorn so many front pages of astronomy
books? What can at first be noticed is that these beautiful galaxies
all belong to a very small and select group; there are simply not many
of such galaxies. What can further be noticed is that these
galaxies generally have companions. Kormendy \& Norman (1979) already
remarked that: ``in general a galaxy shows a global spiral pattern ....
by having a density wave driving mechanism in the form of a bar
or a companion''. For example, well known grand designs with clear
signs of interaction are: M81, M51, M101, NGC 7753, NGC 5364, and
NGC 4305. An interaction with accompanying tidal forcing gives,
in principle, a natural explanation for an emanating $m=2$
symmetry. In addition both theoretical and numerical calculations
(Toomre 1981) show that even a modest external forcing may 
produce a transient grand design spiral pattern. Such a pattern
when viewed from the right angle looks very similar to the observed
morphological and kinematical structure of, for example, M81
(Visser 1980). Considering these arguments and the present
numerical simulations of isolated galaxies, I then conclude
that nearly all non barred grand design spirals are created
by an interaction. 

But how about NGC 628? That galaxy is supposed to be the template
for a typical late type galaxy, yet its observed spiral structure
is more of a grand design nature than can be generated in the
simulations. This apparent contradiction can be resolved by looking
at the neutral hydrogen observations of NGC 628 (Kamphuis \& Briggs
1992) which show that NGC 628 is not that isolated after all.
Instead two gas clouds with masses of $\sim$ 10$^8$ \msol\ are 
situated on opposite sides of the centre at a radius of $\sim$
12\arcmin. In fact the image of these clouds in Fig. 3 of 
Kamphuis \& Briggs is similar to the gravitational forcing
displayed in Fig. 1 of Toomre (1981). That modest forcing already
generates a huge transient spiral wave in the underlying galactic
disc. Although the exact forcing of the gas clouds in NGC 628 is not
known, these clouds can certainly explain
the grand design structure of NGC 628 by swing amplification.

\subsection{On the mass contribution of the disc}
When the stellar
velocity dispersion of a disc is measured the possibilities
for a galaxy's internal constitution are severely restrained. 
For a locally isothermal disc one has the relation
\begin{equation}
\vmd = 0.88\; \dispzn \sqrt{\frac{h}{z_0}},
\end{equation}
Bottema (1993). This equation can be rewritten to 
$z_0 \propto ({\vmd})^{-2}$ and can be drawn onto Fig. 11 once 
the scalelength and dispersion ($\dispzn$) are measured.
For example, lets take NGC 628 being clearly not barred, which restricts
the galaxy to lie in the stable section of Fig. 11.
Unfortunately the stellar velocity dispersion is not measured,
but can be inferred from Fig. 3 of Bottema (1993) such that
$\dispzn = 60 \pm 10$ \kms. For a scalelength of 4.5 kpc, Eq. (33)
when related to Fig. 11 implies that $\vmd <$ 150 \kms, even though
the thickness of the disc is not known. A more massive disc
would bring the galaxy into the bar unstable region, obviously
not consistent with its appearance.

A different approach to determine the contribution of the disc has been
employed by Athanassoula et al. (1987). They use swing amplification
theory to investigate the presence of the amplifier for a certain
mode in case of different disc contributions. For a sample of 48
galaxies with reasonably accurate rotation curves this investigation
has been performed. It appears that, in order to avoid $m=1$,
for the majority of the sample implies that the disc is less massive
than the so called ``maximum disc situation, having a halo without
a hollow core''. In practice that means ${\vmd}/{\vmo} \approx 0.9$.
The threshold for allowing $m=2$ is for the majority of the sample
at an $M/L$ ratio approximately half of that to avoid $m=1$ (not
surprisingly). In terms of rotational values one then has no $m=2$
amplification any more for ${\vmd}/{\vmo} < \frac{1}{2} \sqrt{2}
\times 0.9 = 0.64$. That are the facts presented by Athanassoula et al.
A problem is, however, that it is unclear what is meant with
``requiring that the amplifier is active''. Is a minimum amplification
required? And over what radial extent is it required to be active?
For instance can Fig. 19 be regarded as presenting an active $m=2$ mode?

A second problem is associated with the interpretation.
For example the requirement of no $m=1$ is in contradiction with
the results of the Fourier analysis of Consid\`ere \& Athanassoula
(1988), which shows that for a large fraction of galaxies $m=1$
is present and even dominant. On the other hand it is not known which
fraction of real galaxies indeed has $m=2$ dominant over a certain 
radial extent. Certainly there are a number of galaxies which
do not have $m=2$ present; which are flocculent or amorphous, which
implies that ${\vmd}/{\vmo} < 0.64$ for that category. By using
the swing amplification theory it is assumed that galaxies are
not interacting, such that all $m=2$ activity has an internal origin. 
That is certainly not valid and a number of galaxies could be
featureless when they would be isolated.

Considering the facts presented by Athanassoula et al. and the discussion
just presented, in my opinion, the observed range of morphologies
of galaxies is certainly consistent with the average (non barred) galaxy having
${\vmd}/{\vmo} = 0.64$.

\subsection{On bars}
In Eq. (25) a dimensionless bar stability criterion is given
for a galaxy comparable to NGC 628. The criterion depends on two
parameters: ${\vmd}/{\vmo}$ and the aspect ratio $z_0/h$. Here
I shall elaborate on other parameters that may be of influence for a more 
general galaxy, but shall not go into details on bar formation
(see e.g. Lynden-Bell 1979; Toomre 1981; Binney \& Tremaine 1987, p 381).

Earlier type galaxies have more prominent bulges. Adding a bulge to a disc
plus halo situation has two effects. Firstly,
the rotation curve is made flatter in the inner regions,
creating and ILR. Such an ILR may frustrate or inhibit the formation
of a bar. Secondly adding a bulge means that the disc is dynamically
less important
and the bulge may be considered as
a kind of substitute halo. The combination of both effects results
in a galaxy being more stable against bar formation if its bulge
is more massive. An extreme example of this is presented by
Sellwood \& Evans (2001) where they replace in the inner regions
nearly the complete dark halo by a bulge. The maximum disc rotational
contribution is then 75\% and the minimum $Q$ value amounts to 1.75. That
galaxy is stable, which is not surprising when one considers the
rotational contributions and $Q$ values presented in Sect. 6.

The bar forms just as well with or without the amount of gas 
typical for an Sc galaxy. This is contrary to the claim by 
Shlosman \& Noguchi (1993) that a considerable gas content may
inhibit the formation of a bar. Their calculations, however,
do not include star formation and consequently no SF feedback. As a
result a few ``superclouds'' develop which rapidly spiral inwards
by dynamical friction and strongly heat up the stellar content in
the inner regions. The galaxy is then too hot to form a bar.
In reality star formation sets in and under normal conditions 
such massive gas clouds will never form.

Different bar stability criteria have been derived before. 
At first, for example,
the $T/W$ criterion by Ostriker \& Peebles (1973). That criterion
is derived for MacLaurin spheroids and is as such only applicable
for that kind of theoretical galaxy. Furthermore, the $T$ and $W$
are not easily related to observable quantities. What the calculations
of Ostriker \& Peebles did make very clear was that a large spherical
(dark) matter contribution is an effective way to stabilize against
bars.

A second bar stability criterion put forward by a.o. Sellwood
\& Carlberg (1984) is the requirement of a specified minimum
$Q$ value. The problem is to actually determine that $Q$ value. 
As demonstrated in Sect. 6, $Q$ is not unique, certainly not when
gas is present in a disc. Moreover $Q$ has a radial functionality
which will be different when the radial mass profile is
different. One then has to consider the minimum $Q$ value, or require
$Q$ to be below a certain value over a certain radial extent.
Even then a small difference in $Q$ can mark the change over from
stable to unstable. What can be determined from the investigations
in this paper is that stability problems arise as soon as $Q$ 
($Q_{\ast}$ or $Q_t$) at some radius falls below the value of 
$\sim 1.2$. It may, however, also be 1.1 or 1.3 if the constitution
of the galaxy is slightly different. 

A third criterion is the one by Efstathiou et al. (1982) stating
that a disc is stable to bar formation if
\begin{equation}
\frac{\vmo}{\sqrt{G M_{\rm disc} /h}} \ga 1.1\; .
\end{equation}
This criterion can be rewritten considering that for an exponential
disc $\sqrt{G M_{\rm disc} /h} = 1.61 \vmd$ such that Eq. (34) 
translates into ${\vmd}/{\vmo} \la 0.56$ in order to be stable.
Considering Fig. 11 which contains the criterion of the present study,
it is clear that Efstathiou et al's criterion is overly restrictive.
In addition, for very thin discs with $h/z_0 \ga 12$ it is not valid,
though such thin discs are probably not present in reality.
 
\subsection{Outlook}
It would be desirable to include the effects of dust and metals
in the numerical code. Then matters like FUV shielding, variable
cooling functions, and H$_2$ formation can be investigated.
That might be important in cases like starburst galaxies or ULIRGs.

A grand design in 
a non-barred realistic galaxy still has to be created.
Since that does not appear to be
possible in an isolated galaxy, encounters have to be simulated.
Small satellites passing a large galaxy can be simulated
using a dead halo, but for larger satellites dynamical friction
becomes important and a life halo must be used. Re-creation
of the M51 encounter or the M81-M82 interaction would be
a nice challenge.

An obvious extension of the present work is to investigate
bar stability for galaxies with more massive bulges. A three
parameter stability criterion might then be derived.

It appears difficult to compare numerical spiral structure
with actual spiral structure in a quantitative way. This is
mainly caused by the lack of observational data on spiral structure.
Surely a program to obtain such data is not that difficult. Moreover,
asymmetries in galaxies can then be investigated by determination
of the amount of $m=1$ and from that the presence and frequency
of interactions might be established.

\subsection{Conclusions}
Finally a compilation of the main conclusions following from
the numerical simulations and the subsequent analysis and 
considerations.
\begin{enumerate}
\item
A disc in star formation equilibrium is established by two opposing
actions. On the one hand gas cooling and dissipation, on the other
hand gas heating by the FUV field of young stars and mechanical
SN feedback.
\item
Too little SN feedback results in a disc with giant SF regions;
too much SN feedback disperses the gas and dilutes the spiral
structure.
\item
Spiral features are generated in the disc by shearing gas filements
which swing amplify the underlying mass distribution.
\item
Spiral arms exist for approximately half a rotation period
and disappear by winding and stretching or by interaction with
other spiral features.
\item
A confirmation is given of Ostriker \& Peebles (1973) result that a
late type spiral galaxy is more unstable to bar 
formation when its disc contribution to the total mass is larger.
\item
A late type spiral galaxy is more unstable to bar
formation when its thickness or equivalently its stellar velocity
dispersion is lower.
\item
A bar stability criterion is determined for late type galaxies,
depending on the ratio of maximum disc rotation to total rotation and on
disc thickness.
\item
The bar stability of a disc barely depends on the amount of gas.
\item
At the maximum disc limit a bar stable galaxy can only be created
for a disc which is unrealistically thick.
\item
The simulations display the presence of a substantial 
(20 -- 25\%) amount of luke warm \hi gas in the temperature
range from 1000 to 8000 K.
\item
In the stellar disc the simulated velocity dispersion of the gas
is in good agreement with the observations.
\item
The increase of the stellar velocity dispersion (stellar heating)
of newly formed stars proceeds proportional to $t^{0.5}$, as
observed in the solar neighbourhood. After $\sim$ 6.5 Gyrs the 
velocity dispersion level of the old stellar population is reached.
\item
The simulations generate small and medium scale spiral structure
of which the multiplicity increases as a function of radius. 
\item
It is not possible to create a grand design $m=2$ spiral.
Unless a bar develops which may drive the $m=2$ structure. 
\item
When compared with some exemplary spiral galaxies, the numerically
generated spiral structure is less strong and has a larger
multiplicity.
\item
The emerging spiral structure is qualitatively and quantitatively
consistent with the theory of swing amplification.
\item
Swing amplification explains that for a late type galaxy
a dominant $m=2$ mode over a considerable radial extent
can only exist in a massive disc. Such a disc is unstable
to bar formation. As a consequence a grand design $m=2$ spiral
cannot develop in an isolated bar stable late type galaxy,
just as predicted by the numerical simulations.
\item
The suggestion of Kormendy \& Norman (1979) is confirmed that most
of the grand design $m=2$ spirals must be generated by an interaction
or are driven by a central bar.
\item
The observed range of morphologies of galaxies is consistent
with the average (non barred) galaxy having ${\vmd}/{\vmo} = 0.64$.
\end{enumerate}

\noindent
{\normalsize\bf Acknowledgments}

\noindent
I thank J. Gerritsen for donating his numerical package to me,
for introducing me to its use, and for many discussions.
L. Hernquist is acknowledged for making his TREESPH code available
to the astronomical community. I thank W. de Blok, R. Braun, 
and I. Pelupessy for useful discussions.
The Kapteyn Institute is acknowledged for providing hospitality and support.

\clearpage


\begin{thebibliography}{}
\bibitem[1992]{}
Athanassoula E., 1984, Phys. Rep., 114, 319\par
\bibitem[]{}
Athanassoula E., 1992, MNRAS 259, 345\par
\bibitem[1986]{}
Athanassoula E., Bosma A., Papaioannou S., 1987, A\&A 179, 23\par
\bibitem[]{}
Barbanis B., Woltjer L., 1967, ApJ 150, 461\par
\bibitem[]{}
Barnes J.E., Hernquist L., 1991, ApJ 370, L65\par
\bibitem[]{} 
Barnes J.E., Hut P., 1986, Nature 324, 466\par
\bibitem[1987]{}
Bertin G., Romeo A.B., 1988, A\&A 195, 105\par
\bibitem[]{}
Binney J., Tremaine S., 1987, Galactic Dynamics, Princeton Univ.
Press, Princeton, NJ\par
\bibitem[]{}
Black J.H., 1987, in: Interstellar Processes, Hollenbach D.J.,
Thronson H.A. (eds), Reidel, Dordrecht, p. 731\par
\bibitem[]{}
Bodenheimer P., 1992, in: Star Formation in Stellar Systems,
Tenorio-Tagle G., Prieto M., S\'anchez F. (eds), Cambridge University
Press, p. 3\par
\bibitem[]{}
Bottema R., 1993, A\&A 275, 16\par
\bibitem[]{}
Bottema R., 1996, A\&A 306, 345\par
\bibitem[]{}
Bottema R., 1997, A\&A 328, 517\par
\bibitem[]{}
Bottema R., Gerritsen J.P.E., 1997, MNRAS 290, 585\par
\bibitem[]{}
Braun R., 1995, A\&AS 114, 409\par
\bibitem[]{}
Braun R., 1997, ApJ 484, 637\par
\bibitem[]{}
Bruzual B.A., Charlot S., 1993, ApJ 405, 538\par
\bibitem[]{}
Casertano S., 1983, MNRAS 203, 735\par
\bibitem[]{}
Carlberg R.G., Sellwood J.A., 1985, ApJ 292, 79\par
\bibitem[]{}
Combes F., Becquaert J.-F., 1997, A\&A 326, 554\par
\bibitem[]{}
Consid\`ere S., Athanassoula E., 1982, A\&A 111, 28\par
\bibitem[]{}
Consid\`ere S., Athanassoula E., 1988, A\&AS 76, 365\par
\bibitem[]{} 
Courteau S., Rix H.W., 1999, ApJ 513, 561\par
\bibitem[]{}
Cox D.P., 1990, in: The Interstellar Medium in Galaxies, 
Thronson H.A., Shull J.M. (eds), Kluwer, Dordrecht, p. 181\par
\bibitem[]{}
Dalgarno A., McCray R., 1972, ARA\&A 10, 375\par
\bibitem[]{}
Danver C.G., 1942, Lund Obs. Ann. 10\par
\bibitem[]{}
Debattista V.P., Sellwood J.A., 1998, ApJ 493, L5\par
\bibitem[]{}
de Jong, T., 1977, A\&A 55, 137\par
\bibitem[]{}
Dickey J.M., Brinks E., 1993, ApJ 405, 153\par
\bibitem[]{}
Efstathiou G., Lake G., Negroponte J., 1982, MNRAS 199, 1069\par
\bibitem[]{}
Elmegreen B.G., 1993, ApJ 411, 170\par
\bibitem[]{}
Elmegreen B.G., 1992, in: Star Formation in Stellar Systems,
Tenorio-Tagle G., Prieto M., S\'anchez F. (eds), Cambridge
University Press, p. 383\par
\bibitem[]{}
Elmegreen B.G., Thomasson M., 1993, A\&A 272, 37\par
\bibitem[]{}
Elmegreen B.G., Elmegreen D.M., Seiden P., 1989, ApJ 343, 602\par
\bibitem[]{}
Elmegreen B.G., Elmegreen D.M., Montenegro L., 1992, ApJS 79, 37\par
\bibitem[]{}
Field G.B., Goldsmith D.W., Habing H.J., 1969, ApJ 155, L149\par
\bibitem[]{}
Freeman K.C., 1970, ApJ 160, 811\par
\bibitem[]{}
Friedli D., Benz W., 1993, A\&A 268, 65\par
\bibitem[]{}
Friedli D., Benz W., 1995, A\&A 301, 649\par
\bibitem[]{}
Fuchs B., M\"ollenhoff C., 1999, A\&A 352, L36\par
\bibitem[]{}
Gammie C.F., Ostriker J.P., Jog C.J., 1991, ApJ 378, 565\par
\bibitem[]{}
Gerritsen J.P.E., 1997, Ph. D. Thesis, University of Groningen\par
\bibitem[]{}
Gerritsen J.P.E., de Blok J.W.G., 1999, A\&A 342, 655\par
\bibitem[]{}
Gerritsen J.P.E., Icke V., 1997, A\&A 325, 972 (GI97)\par
\bibitem[]{}
Gingold R.A., Monaghan J.J., 1977, MNRAS 181, 375\par
\bibitem[]{}
Goldreich P., Lynden-Bell D., 1965, MNRAS 130, 125\par
\bibitem[]{}
Habing H.J., 1968, Bull. Astron. Inst. Netherlands 19, 421\par
\bibitem[]{}
Heller C., Shlosman I., 1994, ApJ 424, 84\par
\bibitem[]{}
Hernquist L., 1987, ApJS 64, 715\par
\bibitem[]{}
Hernquist L., Katz N., 1989, ApJS 70, 419\par
\bibitem[]{}
Hollenbach D., Salpeter E.E., 1971, ApJ 163, 155\par
\bibitem[]{}
Huizinga J.E., van Albada T.S., 1992, MNRAS 254, 677\par
\bibitem[]{}
Jog C.J., Ostriker J.P., 1988, ApJ 328, 404\par
\bibitem[]{}
Julian W.H., Toomre A., 1966, ApJ 146, 810\par
\bibitem[]{}
Kalberla P.M.W., Schwarz U.J., Goss W.M., 1985, A\&A 144, 27\par
\bibitem[]{}
Kalnajs A.J., 1975, in: La dynamique des galaxies spirales, 
Colloque International CNRS No. 241, Weliachew L. (ed), p. 103\par
\bibitem[]{}
Kamphuis J.J., 1993, Ph. D. Thesis, University of Groningen\par
\bibitem[]{}
Kamphuis J.J., Briggs F., 1992, A\&A 253, 335\par
\bibitem[]{}
Katz N., 1992, ApJ 391, 502\par
\bibitem[]{}
Katz N., Gunn J.E., 1991, ApJ 377, 365\par
\bibitem[]{}
Kennicutt R.C., 1981, AJ 86, 1847\par
\bibitem[]{}
Kennicutt R.C., 1983, ApJ 272, 54\par
\bibitem[]{}
Kennicutt R.C., 1989, ApJ 344, 685\par
\bibitem[]{}
Kennicutt R.C., 1998, ApJ 498, 541\par
\bibitem[]{}
Kormendy J., Norman C.A., 1979, ApJ 233, 539\par
\bibitem[]{}
Kregel M., van der Kruit P.C., de Grijs R., 2002, MNRAS 334, 646\par
\bibitem[]{}
Kulkarni S.R., Heiles C., 1987, in: Interstellar Processes, Hollenbach D.J.,
Thronson H.A. (eds), Reidel, Dordrecht, p. 87\par
\bibitem[]{}
Lacey C., 1991, in: Sundelius B. (ed), Dynamics of Disc Galaxies,
Dep. of Astronomy/Astrophysics, G\"oteborgs Univ. and Chalmers
Univ. of Technology, G\"oteborg, p. 257\par
\bibitem[]{}
Lamers H.J.G.M.L., Leitherer C., 1993, ApJ 412, 771\par
\bibitem[]{}
Lin C.C., Shu F.H., 1964, ApJ 140, 646\par
\bibitem[]{}
Lin C.C., Shu F.H., 1966, Proc. Nat. Acad. Sci., 55, 229\par
\bibitem[]{}
Lucy L.B., 1977, AJ 82, 1013\par
\bibitem[]{}
Lynden-Bell D., 1979, MNRAS 187, 101\par
\bibitem[]{}
Lynden-Bell D., Ostriker J.P., 1967, MNRAS 136, 293\par
\bibitem[]{}
McCray R., Kafatos M., 1987, ApJ 317, 190\par
\bibitem[]{}
McKee C.F., Ostriker J.P., 1977, ApJ 218, 148\par
\bibitem[]{}
Mihos J.C., Hernquist L., 1994, ApJ 437, 611\par
\bibitem[]{}
Mihos J.C., Hernquist L., 1996, ApJ 464, 641\par
\bibitem[]{}
Monaghan J.J., 1989, J. of Comp. Physics 82, 1\par
\bibitem[]{}
Mueller M.W., Arnett W.D., 1976, ApJ 210, 670\par
\bibitem[]{}
Natali G., Pedichini F., Righini M., 1992, A\&A 256, 79\par
\bibitem[]{}
Navarro J.F., White S.D.M., 1993, MNRAS 265, 271\par
\bibitem[]{}
Ostriker J.P., Peebles P.J.E., 1973, ApJ 186, 467\par
\bibitem[]{}
Payne H.E., Salpeter E.E., Terzian Y., 1983, ApJ 272, 540\par
\bibitem[]{}
Rix H.W., Zaritsky D., 1995, ApJ 447, 82\par
\bibitem[]{}
Romeo A.B., 1994, A\&A 286, 799\par
\bibitem[]{}
Rubin V.C., Burstein D., Kent Ford W., Thonnard N., 1985,
ApJ 289, 81\par
\bibitem[]{}
Ryder S.D., Dopita M.A., 1994, ApJ 430, 142\par
\bibitem[]{}
Salucci P., Ashman K.M., Persic M., 1991, ApJ 379, 89\par
\bibitem[]{}
Sandage A., Tammann G.A., 1981, A Revised Shapley-Ames Catalog
of Bright Galaxies, Carnegie Institute of Washington\par
\bibitem[]{}
Schmidt M., 1959, ApJ 129, 243\par
\bibitem[]{}
Sellwood J.A., Carlberg R.G., 1984, ApJ 282, 61\par
\bibitem[]{}
Sellwood J.A., Evans N.W., 2001, ApJ 546, 176\par
\bibitem[]{}
Sellwood J.A., Moore E.M., 1999, ApJ 510, 125\par
\bibitem[]{}
Shlosman I., Noguchi M., 1993, ApJ 414, 474\par
\bibitem[]{}
Shostak G.S., van der Kruit P.C., 1984, A\&A 132, 20\par
\bibitem[]{}
Shu F.H. Adams F.C., Lizano S., 1987, ARA\&A 25, 23\par
\bibitem[]{}
Silich S.A., Franco J., Palou\u{s} J., Tenorio-Tagle G.,
1996, ApJ 468, 722\par
\bibitem[]{}
Spitzer L., 1978, Physical Processes in the Interstellar Medium,
Wiley \& Sons, New York, p. 142\par
\bibitem[]{}
Spitzer L., Schwarzschild M., 1953, ApJ 118, 106\par
\bibitem[]{}
Springel V., 2000, MNRAS 312, 859\par
\bibitem[]{}
Toomre A., 1964, ApJ 139, 1217\par
\bibitem[]{}
Toomre A., 1981, in: The Structure and Evolution of Normal Galaxies,
eds. Fall S.M. and Lynden-Bell D., Cambridge Univ. Press, Cambridge, p.111\par
\bibitem[]{}
Tremaine S., Ostriker J.P., 1999, MNRAS 306, 662\par
\bibitem[]{}
van Albada T.S., Sancisi R., 1986, Phil. Trans. R. Soc. London, Ser. A,
320, 447\par
\bibitem[]{}
van der Hulst J.M., 1996, in: The Minnesota lectures on extragalactic
neutral hydrogen, Astronomical Society of the Pacific Conference Series,
Vol. 106, Skillman E.D. (ed), p. 47\par
\bibitem[]{}
van der Kruit P.C., 1995, in: van der Kruit P.C., Gilmore G. (eds) IAU
Symposium No. 164, Stellar Populations, Kluwer, Dordrecht, p. 205\par
\bibitem[]{}
van der Kruit P.C., Searle L., 1981, A\&A 95, 105\par
\bibitem[]{}
van der Kruit P.C., Searle L., 1982, A\&A 110, 61\par
\bibitem[]{}
Villumsen J.V., 1985, ApJ 290, 75\par
\bibitem[]{}
Visser H.C.D., 1980, A\&A 88, 149\par
\bibitem[]{}
Wevers B.M.H.R., van der Kruit P.C., Allen R.J., 1986, A\&AS 66, 505\par
\bibitem[]{}
Wielen R., 1977, A\&A 60, 263\par
\bibitem[]{}
Wolfire M.G., Hollenbach D., McKee C.F., et al., 1995, ApJ 443, 152\par
\bibitem[]{}
Young L.M., Lo K.Y., 1996, ApJ 462, 203\par
\end{thebibliography}
\end{document}